\documentclass[a4paper]{amsart}
\usepackage{amsmath,amssymb, amscd}
\usepackage{pdfsync}
\usepackage{graphicx,a4wide}
\usepackage{cite}
\usepackage{url}

\newtheorem{theorem}{Theorem}[section]

\newtheorem{remark}[theorem]{Remark}

\newcommand{\I}{\mathrm{i}}
\newcommand{\D}{\mathrm{d}}
\newcommand{\name}[1]{\texttt{#1}}

\begin{document}

\title{Computational approach to compact Riemann surfaces}

\author[J.~Frauendiener]{J\"org Frauendiener}
\address[J.~Frauendiener]{Department of Mathematics and Statistics, University of Otago,     
P.O. Box 56, Dunedin 9010, New Zealand}
\email{joergf@maths.otago.ac.nz}
\author[C.~Klein]{Christian Klein}
 \address[C.~Klein]{Institut de 
Math\'ematiques de Bourgogne,
Universit\'e de Bourgogne,
9 avenue Alain Savary,
BP 47970, 21078 Dijon Cedex,
France}
\email{christian.klein@u-bourgogne.fr}
\thanks{
This work was supported in part by the Marie Curie IRSES program 
RIMMP.  JF thanks for the hospitality at the University of Burgundy as a 
visiting professor, where part of this work has been completed. 
}
\begin{abstract}
  A purely numerical approach to compact Riemann surfaces starting from plane
  algebraic curves is presented. The critical points of the algebraic curve are
  computed via a two-dimensional Newton iteration. The starting values for this
  iteration are obtained from the resultants with respect to both coordinates of
  the algebraic curve and a suitable pairing of their zeros.  A set of
  generators of the fundamental group for the complement of these critical
  points in the complex plane is constructed from circles around these points and
  connecting lines obtained from a minimal spanning tree. The monodromies are
  computed by solving the defining equation of the algebraic curve on
  collocation points along these contours and by analytically continuing the
  roots. The collocation points are chosen to correspond to Chebychev
  collocation points for an ensuing Clenshaw-Curtis integration of the
  holomorphic differentials which gives the periods of the Riemann surface with
  spectral accuracy.  At the singularities of the algebraic curve, Puiseux
  expansions computed by contour integration on the circles around the
  singularities are used to identify the holomorphic differentials.  The Abel
  map is also computed with the Clenshaw-Curtis algorithm and contour
  integrals. As an application of the code, solutions to the
  Kadomtsev-Petviashvili equation are computed on non-hyperelliptic Riemann
  surfaces.
    
\end{abstract}
\maketitle

\section{Introduction}
\subsection{Background and Motivation}

Riemann surfaces are important in many fields of mathematics and physics. An
interesting example for their applications are quasiperiodic solutions to
certain integrable partial differential equations (PDEs) such as Korteweg-de
Vries (KdV) and nonlinear Schr\"odinger (NLS) equation given at the beginning of
the 1970s by Novikov, Dubrovin, Its, Matveev, van Moerbeke, Krichever and others
in terms of multidimensional theta functions on compact Riemann surfaces, see
\cite{algebro,Dintro} and references therein for a historic account.  Whereas
the surfaces are hyperelliptic in the case of the KdV and NLS equation,
Krichever's \cite{K} solutions to the Kadomtsev-Petviashvili (KP) equation can
be given on arbitrary compact Riemann surfaces.

These applications require efficient numerical tools to 
study Riemann surfaces in cases where analytical methods are not yet 
available. First plots of KP solutions appeared in \cite{Mum} and via 
Schottky uniformizations in \cite{Bob}. Since all compact Riemann 
surfaces can be defined via non-singular plane algebraic curves of the form
\begin{equation}
    f(x,y) = 
    \sum_{i=1}^{M}\sum_{j=1}^{N}a_{ij}x^{i}y^{j}=\sum_{j=1}^{N}a_{j}(x)y^{j}
    \label{fdef}\;
\end{equation}
with constant complex coefficients $a_{nm}$, Deconinck and van Hoeij developed
an approach to the symbolic-numerical treatment of algebraic curves. Their work
is distributed as the \name{algcurves} package with Maple, see
\cite{deco01,deconinck03,RSbookdp}. It works for coefficients given as exact
arithmetic expressions, a restriction due to the use of exact integer
arithmetic. A purely numerical approach to real hyperelliptic Riemann surfaces
was given in \cite{prd,cam,lmp} and for general hyperelliptic 
surfaces in \cite{hyper}.  For a review of the current state of the
art of computational approaches to Riemann surfaces, the reader is referred to
\cite{Bob}.  A first adaption of the \name{algcurves} package to Matlab for
general complex coefficients $a_{nm}$ was presented in \cite{alg1}.  The fact
that the coefficients are general floating point numbers, removed the
limitations in the Maple package on the coefficients which have to be converted
to rational numbers, a procedure which can lead to technical problems in
practice especially when families of algebraic curves are studied, i.e., when
the coefficients $a_{nm}$ in (\ref{fdef}) themselves depend on parameters and
when one is interested in modular properties of Riemann surfaces. An additional
problem in this context can be computing time since the computation of the
Riemann matrix uses the somewhat slow Maple integration routine. Thus, a more
efficient computation of the Riemann matrix is interesting if one wants to study
families of Riemann surfaces or higher genus examples which are computationally
expensive. Therefore, the numerical approach to compact Riemann surfaces
presented in this paper is complementary to the Maple \name{algcurves} package:
whereas for examples of curves of low genus with rational coefficients the Maple
approach is excellent, the purely numerical approach is intended for the study
of modular properties and computationally demanding higher genus surfaces. Note,
that a transfer of the Maple package to a more efficient platform is also
planned by the original authors of the \name{algcurves} package, see \cite{sage}
for a first step in this direction.

The numerical approach presented in this paper allows to efficiently study
solutions to integrable PDEs along the lines discussed in \cite{KK1} and, for
real nonsingular solutions, in \cite{KK2}. As an example we will consider here
solutions to the KP equation on non-hyperelliptic Riemann surfaces. The code
can also be used to efficiently study modular properties of Riemann surfaces, which
have many applications. Examples in this context are determinants of Laplacians
on Riemann surfaces which appear for instance in conformal field theories, see
\cite{Sarnak} for an overview.  A numerical study of extremal points for the
determinant of the Laplacian in the Bergman metric on Riemann surfaces of genus
2 was presented in \cite{klkoko}.  The modular dependence of Riemann surfaces is
also important in the asymptotic description of so-called dispersive shocks,
highly oscillatory regions in solutions to purely dispersive equations such as
KdV and NLS.  The asymptotic description of the rapidly modulated oscillations
in the dispersive shock is given by the exact solution to the KdV equation on a
family of elliptic or hyperelliptic surfaces, where, however, the branch points depend on the
physical coordinates via the Whitham equations \cite{whitham1,W}, see
\cite{LL,V1,DVZ,GK,kam2,to1}.  Families of hyperelliptic curves, where the
branch points depend on the physical coordinates, also appear in exact solutions
for the Ernst equation, which has many applications in mathematics such as the
theory of Bianchi surfaces, and physics such as general relativity, see
\cite{ernstbook} for references. Hyperelliptic solutions to the Ernst equation
were found by Korotkin \cite{korot1}.  The Einstein-Maxwell equations in the
presence of two commuting Killing vectors are equivalent to the electro-magnetic
Ernst equations. The latter have solutions on three-sheeted coverings of
$\mathbb{C}P^{1}$ with branch points depending on the physical coordinates, see
\cite{korot1,annalen}.

\subsection{Main numerical approaches}

The preliminary version of the first fully numerical approach to Riemann
surfaces in \cite{alg1} was very close to the \name{algcurves} package. It was
found that the methods used in that package sometimes led to \emph{cancellation
  problems} near singularities of the curves where numerically large terms
cancel and thus lead to a comparatively high inaccuracy in the final result, a
well known problem for numerical approaches. In this paper we present a
computational approach to algebraic curves which is better suited for a purely
numerical treatment. The critical points of the algebraic curve are identified
with a two-dimensional Newton iteration, the paths for the computation of the
monodromies are constructed in numerically optimal form as in \cite{FKS} (see
also \cite{poteauxj}) with a minimal spanning tree. The periods of the
holomorphic differentials are computed with the Clenshaw-Curtis method which
shows an exponential decrease of the numerical error with the number of
collocation points for the analytic functions integrated here. This is to be
contrasted with widely used finite difference techniques for which the numerical
error decreases algebraically with the number of collocation points.

A general problem in finite precision numerical approaches are
\emph{cancellation errors} which occur when numerically large terms cancel
resulting in a small residue with a comparatively large numerical error (since
all numerical results are given with the same finite precision). To avoid such
errors, we make use at several occasions of the well known Cauchy formula
\begin{equation}
    \mathcal{F}(t) = \frac{1}{2\pi \mathrm{i}}\int_{\Gamma}^{}\frac{\mathcal{F}(t')dt'}{t'-t}
    \label{cauchy}.
\end{equation}
Here $\mathcal{F}$ could be a holomophic differential or the Abel map, and
$\Gamma$ is some closed contour on the Riemann surface around the point $t$ at
which cancellations occur in the evaluation of the quantity $\mathcal{F}$. Since
we are interested in the evaluation of such quantities at singularities or
branch points of the algebraic curve, the integrand will be already known on the
contours used to compute the periods, and the computation of the integral in
(\ref{cauchy}) comes at essentially no additional cost. In addition there will
be no cancellations on the contours $\Gamma$ since by construction they avoid
the critical points of the algebraic curve.  Thus, machine precision ($10^{-16}$
in our case, in practice limited to $10^{-12}$ because of rounding errors) can
be reached also in cases where this was not possible in \cite{alg1}.

The paper is organized as follows: In Sect.~\ref{2} we determine the critical
points of an algebraic curve, i.e., branch points and singular points, via a
two-dimensional Newton iteration. The starting points of the iteration are
obtained from the zeros of the resultants of $f$ and $f_{y}$ with respect to
both coordinates. The paths for the computation of the monodromies, the
generators of the fundamental group of the complex plane minus the critical
points of the algebraic curve, are constructed in Sect.~\ref{4} from circles
around these points and connecting lines according to a minimal spanning
tree. The algebraic equation for the curve is solved at a set of points on these
contours.  In view of the numerical integration, these collocation 
points are chosen as
\emph{Chebychev points}.  Then the computation of the integrals of the candidates for
the holomorphic differentials is only a scalar multiplication at negligible
computational cost.  In Sect.~\ref{3} we compute the Puiseux expansions at the
singular points via the Cauchy formula.  These expansions are used in
Sect.~\ref{3a} to determine a basis of the holomorphic 1-forms.  The periods
of the holomorphic 1-forms are computed for a homology basis constructed with
the Tretkoff and Tretkoff algorithm \cite{tret}.  The Abel map is computed for
general points on the Riemann surface in Sect.~\ref{9a}.  In Sect.~\ref{7} we
discuss the numerical performance and convergence properties in dependence of
the numerical resolution in the integration.  In Sect.~\ref{9} we use the
characteristic quantities of the Riemann surface such as the Riemann matrix
obtained above to compute multi-dimensional theta functions. This allows for an
efficient computation of solutions to certain completely integrable equations as
KP, which we discuss as an example in Sect.~\ref{10}.  In sect.~\ref{outlook} we
add some concluding remarks.

\section{Branch points and singular points}
\label{2}
In this section, we numerically identify the critical points of the algebraic
curve (\ref{fdef}) for a projection on the $x$-plane, i.e., the branch points
and singular points of the curve given by the set of common zeros of $f(x,y)$
and $f_{y}(x,y)$, the partial derivative of $f$ with respect to $y$. In
\cite{alg1}, these points were determined via the resultant of $f$ and $f_{y}$,
and the zeros of the resulting polynomial in $x$ were identified via Zeng's
\cite{zeng} \name{multroot} package in Matlab. Here we use the resultant (both
for a projection on $x$ and on $y$) only to obtain starting values for an
ensuing two-dimensional Newton iteration with some post processing for multiple
zeros.

\subsection{Starting values for the iteration}

In this paper, we consider plane algebraic curves defined as a subset $C$ of
$\mathbb{C}^{2}$ where $C = \{(x,y)\in\mathbb{C}^{2}|f(x,y)=0\}$, where $f(x,y)$
is an irreducible polynomial in $x$ and $y$ of the form (\ref{fdef}).  We assume
that not all $a_{iN}$ vanish and that $N$ is thus the degree of the polynomial
in $y$. The joint degree in $x$ and $y$, i.e., the maximum of $i+j$ for
non-vanishing $a_{ij}$ is denoted by $d$.  We will always solve (\ref{fdef}) for
$y$. In general position, for each $x$ there are $N$ distinct solutions $y_{n}$
corresponding to the $N$ sheets of the Riemann surface. At the points where
$f_{y}(x,y)$ vanishes, there are less than $N$ distinct solutions and thus less
than $N$ sheets. These points are either branch points or singularities. The $x$
coordinates of the points where $f(x,y)=0$ and $f_{y}(x,y)=0$ are given by the
zeros of the resultant $R(x)$ of $Nf-f_{y}y$ and $f_{y}$, the discriminant of
the curve. The resultant is given in terms of the $2N\times 2N$ Sylvester
determinant,
\begin{equation}
    \begin{split}
   & R(x) =    \label{resultant}\\
   & \begin{pmatrix}
        a_{N-1} & 2a_{N-2} & \ldots & Na_{0} & 0&\ldots & \ldots&0  \\
	0&a_{N-1} & 2a_{N-2} & \ldots & Na_{0} & 0&\ldots & 0   \\
        \vdots & \ddots &  &  & & & \ddots&\vdots  \\
	0&\ldots & \ldots& 0& a_{N-1} & 2a_{N-2} & \ldots & Na_{0}   \\
	Na_{N-1} & (N-1)a_{N-2} & \ldots & a_{1} & 0&\ldots & \ldots&0  \\
	0&Na_{N-1} & (N-1)a_{N-2} & \ldots & a_{1} & 0&\ldots & 0 \\
	\vdots & \ddots &  &  & & & \ddots&\vdots  \\
	0&\ldots & \ldots&0 &Na_{N-1} & (N-1)a_{N-2} & \ldots & a_{1}
    \end{pmatrix}
    \end{split}.
\end{equation}

The algebraic curve is completely characterized by the matrix $a_{ij}$
in (\ref{fdef}). Each entry in the Sylvester determinant is one of the
functions $a_{j}(x)=\sum_{i=1}^{M}a_{ij}x^{i}$ depending on $x$ and thus by
itself a vector of length $M$.  Therefore, the computation of the determinant
involves products of the $a_{i}$, i.e., convolutions of these vectors which are known to be
equivalent to products in Fourier space.  To compute the resultant, we
build as in \cite{alg1} the Sylvester determinant (\ref{resultant})
of the discrete Fourier transforms of the vectors
$a_{n}=(a_{1n},\ldots,a_{Mn})^{T}$.  Each vector in this determinant
is divided by $N$ for numerical reasons.  The determinant is obtained
in Fourier space, and the resultant follows from this via an inverse
Fourier transform. An analogous resultant can be constructed for the 
projection on $y$ leading to a polynomial in $y$ only.

The roots of the resulting polynomial in $x$ gives the $x$ coordinates of the
points, where $f(x,y)=f_{y}(x,y)=0$. Note that the resultant is also used by the
\name{algcurves} package to determine the critical points. The use of exact
integer arithmetics in that package has the consequence that there are no
rounding problems. This is in contrast to the numerical approach of \cite{alg1}
and the present paper, where finite precision arithmetic is used. This
necessitates a careful rounding of all numerical results to a certain number of
digits (limited by machine precision). Typically we aim at a
precision $\mathtt{Tol}$ which can be freely chosen between $10^{-10}$ and
$10^{-14}$.

Root finding in Matlab is possible via the \name{roots} function. It
uses efficient algorithms to find the eigenvalues of the
\emph{companion matrix}, i.e., the matrix which has the studied polynomial as
the characteristic polynomial. The eigenvalues are determined to
machine precision which does not mean, however, that the zeros of the
polynomial with coefficients within roundoff error are determined with
machine precision. Problems occur if there are multiple roots or roots
which are almost identical. It is well known that the computation of
multiple roots is a long standing numerical challenge, see for
instance \cite{zeng} for references. The most common approaches in
this context use multiprecision arithmetic, i.e., more than 16 digits,
and need exact coefficients of the polynomials.  This is also the 
main reason why the Maple \name{algcurves} package uses exact integer 
arithmetic. However, if the
coefficients of the polynomials are not exact, but approximated by 
floating point numbers, this will inhibit the
identification of correct multiple roots. Finite precision in the
coefficients of the polynomial turns multiple roots into clusters of
simple roots. Consider for example the case of the Klein curve, the
curve of lowest genus with the maximal number of automorphisms, in the
form
\begin{equation}
    y^{7}=x(x-1)^{2}
    \label{klein}\;.
\end{equation}
The resultant with respect to $x$ for this curve has the form
$R(x)=x^{6}(x-1)^{12}$. After rounding, our procedure gives the
correct coefficients of the polynomial up to machine precision, but
instead of the root at 1 with multiplicity 12, \texttt{roots(R(x))}
returns the following cluster of roots,
\begin{verbatim}        
       1.1053 + 0.0297i
       1.1053 - 0.0297i
       1.0736 + 0.0790i
       1.0736 - 0.0790i
       1.0224 + 0.1032i
       1.0224 - 0.1032i
       0.9686 + 0.0980i
       0.9686 - 0.0980i
       0.9264 + 0.0686i
       0.9264 - 0.0686i
       0.9037 + 0.0245i
       0.9037 - 0.0245i,
\end{verbatim}
a consequence of the finite precision (the error in the above roots is of the order of 
the 12th root of the rounding error).

The identification of multiple roots within the available numerical precision is
thus a crucial problem in the numerical study of algebraic curves with non-exact
coefficients. In \cite{alg1} we used Zeng's Matlab package \name{multroot}.  As
discussed in more detail in \cite{zeng}, two algorithms are used by
\name{multroot} to identify multiple roots: The first algorithm identifies
tentatively the multiplicity of the roots, the second uses a Newton iteration to
determine the roots corresponding to this multiplicity structure to machine
precision. The code provides an estimation of the forward and backward
error\footnote{As usual the forward error for the approximation of the value of
  a function $k(x)$ at some given point $x$ via an approximate function
  $\tilde{k}(x)$ is defined as the difference $k(x)-\tilde{k}(x)$; the backward
  error is defined as the difference $\tilde{x}-x$, where $\tilde{x}$ is the
  value for which $k(\tilde{x})=\tilde{k}(x)$.} and varies with the multiplicity
structure to minimize the backward error.  The \texttt{multroot} package is very
efficient. For the above example of the Klein curve (\ref{klein}), it finds the
two zeros 0 and 1 with multiplicity 6 and 12, respectively. As mentioned above,
rounding is important in this context.

\subsection{Two-dimensional Newton iteration}

The disadvantage of the approach to identify the critical points of an algebraic
curve via the resultant is that the zeros of a high order polynomial in one
variable have to identified. This gives the values of $x$, and the corresponding
values of $y$ are computed a posteriori by solving (\ref{fdef}) for each of
these $x$ values. If the degree of the polynomial obtained from the resultant
gets too high, the ratio of the estimated forward to backward error will be
comparable to the maximal precision accessible in Matlab, and the results for
the roots via \name{multroot} will not be reliable.  Instead we solve here the
equations for both $x$ and $y$ values simultaneously, i.e., we solve two
polynomial equations in two variables at the same time instead of one polynomial
in one variable of considerably higher order than $N$. In the example of the
Klein curve (\ref{klein}), these would be the polynomials $x(x-1)^{2}$ and
$y^{6}$. Visibly the multiplicity of the zero in $x$ at $x-1$ is much lower than
for the resultant, 2 instead of 12. Whereas the problems with multiple zeros
persist, they are thus of considerably lower order.

If, as in the example of the Klein curve or of hyperelliptic curves, one of the
polynomials $Nf-yf_{y}$ or $f_{y}$ depends only on one variable $x$ or $y$, the
zeros of this polynomial are determined with \name{roots}. The resulting roots
are taken as starting values for a standard one-dimensional Newton iteration to
improve the accuracy of the found roots and to identify multiple roots as
discussed below for the general case. The second polynomial is then used to
determine for each of these roots the corresponding values of the second
variable $x$ or $y$.

In the general case where none of the two polynomial equations to be solved
reduces to one dimension, the starting values for the Newton iteration are
obtained from the resultants of $Nf-yf_{y}$ and $f_{y}$ with respect to both $x$
and $y$. The zeros $x_{k}$, $k=1,\ldots,N_{x}$ and $y_{i}$, $i=1,\ldots,N_{y}$
(since some of the finite zeros in one variable can correspond to infinite
values in the other, $N_{x}$ is not necessarily equal to $N_{y}$), of these
polynomials are identified via \name{roots}.  As in the example of the Klein
curve above, multiple roots are determined with a considerable numerical
error. Therefore the roots need to be paired in a way to minimize the residual
of the two equations $Nf-yf_{y}$ and $f_{y}$. Note that this residual can be
rather large for zeros of high multiplicity. Up to values of $10^{6}$ for this
quantity, we observe convergence of the Newton iteration up to machine
precision, in general within a few iterations.

Whereas the found roots will satisfy the equations $Nf-yf_{y}=0$ and $f_{y}=0$
to machine precision, as in the example of the Klein curve this does not mean
that multiple roots are identified with this precision. Thus roots which
coincide to a certain precision, which can be chosen by hand, are identified.
In the examples discussed in the following, a tolerance of $10^{-6}$ was
chosen. If the ratio of the smallest distance between any two critical points to
the largest distance between any two finite critical points is smaller than
$10^{-3}$, the code issues a warning.

Given the multiplicity of the roots of the equations $Nf-yf_{y}=0$ and
$f_{y}=0$, the code determines the singular points, i.e., the points with
$f(x,y)=f_{x}(x,y)=f_{y}(x,y)=0$.  All roots $(x_{s},y_{s})$ with a multiplicity
greater than one are tested in this context: it is checked whether
$f_{x}(x_{s},y_{s})=0$ with the same precision used for the check of multiple
roots.  In this way we find all finite branch points and singularities.

\subsection{Singularities at infinite points}

To determine the singular behavior of the curve $f(x,y)=0$ at infinity, we
proceed similarly as the Maple package: we introduce homogeneous coordinates
$X,Y,Z$ via $x=X/Z$, $y=Y/Z$ in (\ref{fdef}) and get
\begin{equation}
    F(X,Y,Z) := 
    Z^{d}f(X/Z,Y/Z)=\sum_{i=1}^{M}\sum_{j=1}^{N}a_{ij}X^{i}Y^{j}Z^{d-i-j}=0
    \label{Fdef}\;.
\end{equation}
Infinite points of the algebraic curve are given by $Z=0$, for the finite points
one can choose $Z=1$. Singular points at infinity satisfy
$F_{X}(X,Y,0)=F_{Y}(X,Y,0)=F_{Z}(X,Y,0)=0$. We first check for such points with
$Y\neq 0$ which implies we can put $Y=1$ without loss of generality. The roots
of $F_{X}(X,1,0)=0$ are determined via \name{roots} and an ensuing Newton
iteration.  It is then checked as above whether the found roots also satisfy
$F_{Y}(X,1,0)=0$ and $F_{Z}(X,1,0)=0$. This analysis identifies all singular
points with $Y\neq 0$, but not the ones with $Y=0$ and $X\neq0$.  In the latter
case we can put $X=1$ and check whether
$F_{X}(1,0,0)=F_{Y}(1,0,0)=F_{Z}(1,0,0)=0$. The singularities are given by the
code in homogeneous coordinates in the form $\mbox{sing}=[X_{s},Y_{s},Z_{s}]$.

\subsection{Example}

As in \cite{RSbookdp,alg1} we will consider the curve
\begin{equation}
  f(x,y)=y^{3}+2x^{3}y-x^{7}=0\;,
  \label{cubic}
\end{equation} 
throughout as an example for the various aspects of the code, if necessary
complemented by further curves. For (\ref{cubic}) we find the finite branch
points (corresponding to 0 and $-2/3^{3/5}$ times fifth roots of
unity)\footnote{For the ease of representation we only give 4 digits here even though
  Matlab works internally with 16 digits.}
\begin{verbatim}
bpoints =
  -0.3197 - 0.9839i
   0.8370 - 0.6081i
   0.8370 + 0.6081i
  -0.3197 + 0.9839i
  -0.0000 + 0.0000i
  -1.0346 + 0.0000i
\end{verbatim}
and  two singularities,
\begin{verbatim}
sing =
  0     0     1     4
  0     1     0     9
 \end{verbatim}
corresponding to $x=y=0$ and $Y=1$, $X=Z=0$. The last column corresponds 
to the delta invariant at the respective singularity, for a 
definition of which we 
refer to  Sect.~\ref{3a}. The fact that the point $x=0$ appears both 
as a branch point and as a singularity implies that it is a 
discriminant point which is in addition singular.

\section{Paths for the computation of the monodromies}%
\label{4}

In this section we describe how to construct the paths for the computation of
the monodromies. As in the Maple \name{algcurves} package and in
\cite{alg1,FKS}, we consider the \emph{problem
  points} determined in the previous section, i.e., the branch points and
singularities of a given algebraic curve and the finite points where $y$
diverges (the zeros of $a_{N}(x)$ determined with the same techniques as
explained in the previous section), which are denoted by $b_{i}$,
$i=1,\ldots,N_{c}$.  The paths to compute the monodromies 
are formed by a set of circles around these problem points 
and by connecting lines between these circles. As in
\cite{FKS} and \cite{potthese}, these connecting lines will be formed according
to a minimal spanning tree.  In contrast to \cite{deco01,alg1,FKS}, we do not
consider half circles around the discriminant points obtained by intersecting
the circles with lines through the points $b_{i}$, $i=1,\ldots,N_{c}$, parallel
to the real axis. In that approach, the lines connecting the circles always at
these intersection points. In the present approach, we use essentially direct
connections between the circles since this has numerical advantages. As in
\cite{FKS}, we first construct the paths $\gamma_{i}$, $i=1,\ldots,N_{c}$, then
we determine how the numbering of the points must be changed by swapping points
in order to ensure the relation
\begin{equation}
    \gamma_{1}\gamma_{2}\cdots 
    \gamma_{N_{c}}\gamma_{\infty}=\mathrm{id}
    \label{gamma},
\end{equation}
where $\gamma_{\infty}$ is the contour surrounding all $b_{i}$, 
$i=1,\ldots,N_{c}$ in clockwise order. 


To obtain a set of generators for the fundamental group
$\pi_1(\mathbb{C}P^1\setminus{\{b_i\}_{i=1}^{N_c}})$, we construct a set of
contours $\gamma_{i}$, $i=1,\ldots,N_{c}$, i.e., closed paths in the base of the
covering starting at the base point $a$ (chosen as explained below) and going
around each of the $b_{i}$, $i=1,\ldots,N_{c}$, projected into the base.
Numerical problems are to be expected if a contour $\gamma_{i}$ comes too close
to one of the problem points $b_{j}$, $j=1,\ldots,N_{c}$. Thus, it is necessary
in the construction of the contours that all of them have a minimal distance
from all problem points.

As in \cite{RSbookdp,alg1}, we determine the minimal distance $\rho$ between 
any two problem points,
\begin{equation*}
    \rho := \min\limits_{
    \begin{smallmatrix}
        i,j=1,\ldots,N_{c}\\
        i\neq j
    \end{smallmatrix}}
   (|b_{i}-b_{j}|)\;.
\end{equation*}
We choose a radius $R=\kappa \rho$, where $0<\kappa<0.5$ for circles 
around these discriminant points. For a better 
vectorization\footnote{This denotes the simultaneous execution of 
similar commands by a computer.} of the 
code we use the same value of $R$ for all discriminant points in 
contrast to the Maple \name{algcurves} package, where such a radius is 
determined for each of them. We typically work with values of 
$\kappa$ between $1/3$ and $1/2.1$ (in Maple, the value $\kappa=2/5$ 
is used).

\begin{figure}[htb]
    \centering 
    \includegraphics[width=10cm]{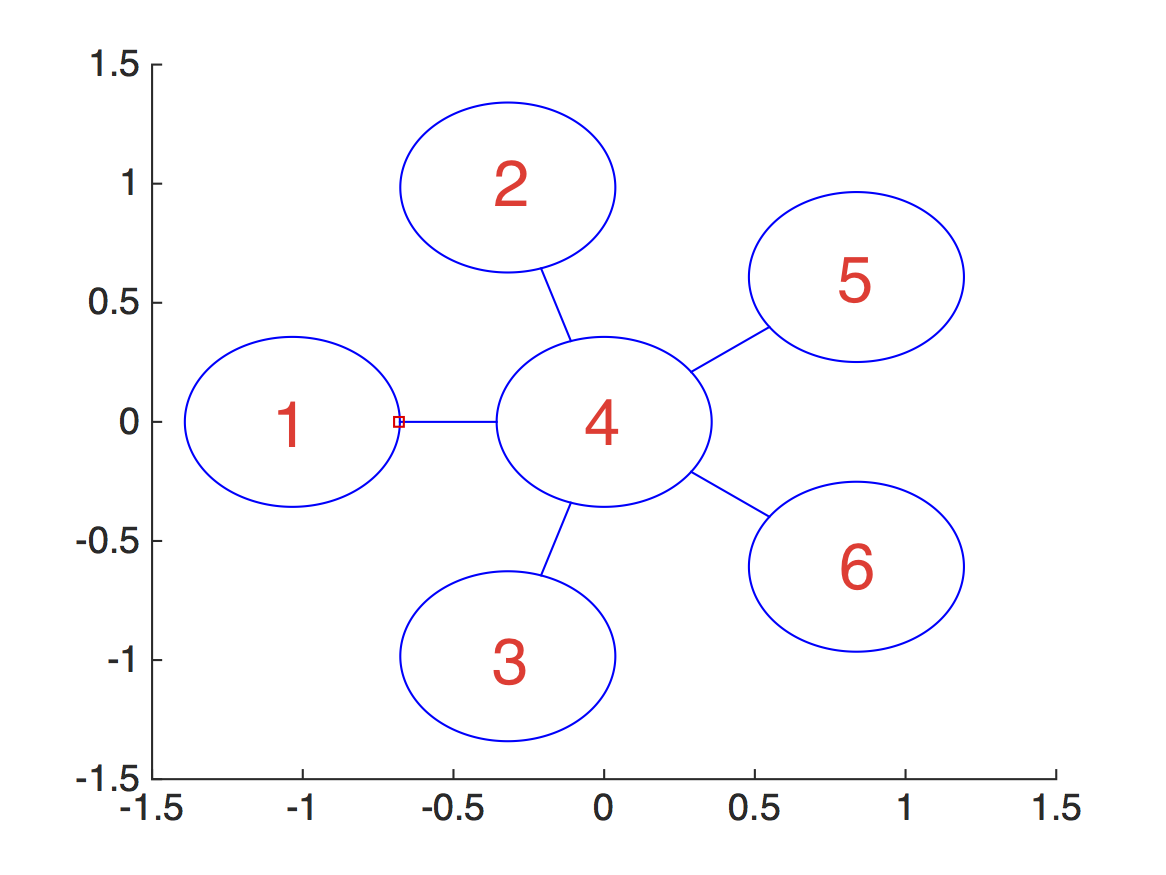}
    \caption{Paths for the computation of the monodromies for the curve 
    (\ref{cubic}) with a radius of the circles around the 
    discriminant points $R=\rho/2.9$, where $\rho$ is the minimal 
    distance between any two branch points. The base point is marked with 
    a square.}
    \label{fig:monpaths}
\end{figure}

The general procedure to construct the contours $\gamma_i$ is as follows: The
points are ordered according to a minimal spanning tree, starting with the point
with minimal real part (in the case of a degeneracy, the point with minimal
imaginary part amongst those with minimal real part is chosen). Then the problem
point nearest to this starting point is identified, after that the point nearest
to one of these two points and so on. As indicated by the so constructed minimal
spanning, direct connection lines are drawn between neighboring points.  The
base point $a$ is chosen as the intersection point of the line through the first
two problem points with the circle around the first point. For the example of the
curve (\ref{cubic}), this leads to Fig.~\ref{fig:monpaths}, where the base point
is indicated with a square. The part of the line through the first two points
(here denoted by 1 and 4) between the circles around the two points forms the
connecting line between the circles around the points $1$ and $4$.  In the same
way connecting lines between all problem points in the order indicated by the
spanning tree are constructed which leads to Fig.~\ref{fig:monpaths}. In
practice a set of $N_{l}+1$ collocation points (as detailed in the next section)
is introduced on each of the circles starting with the first connecting
line to this circle (in the order indicated by the spanning tree) and going
around the circle in counter-clockwise order. All further connecting lines from
a circle to a neighboring circle then start at the collocation point
closest to the direct connection between these two points. The contours
$\gamma_{i}$, $i=1,\ldots,N_{c}$ are then formed by the circle around the point
$b_{i}$, connecting lines between circles and segments of circles around other
points. In Fig.~\ref{fig:monpaths}, the contour $\gamma_{3}$ for instance starts
at the base point, goes along the connecting line between $b_{1}$ and $b_{4}$,
then in counter-clockwise order to the connecting line between $b_{4}$ and
$b_{3}$, along the circle around $b_{3}$ and then the same way back to the base
point.

As already mentioned, in the Maple \name{algcurves} package and in
\cite{alg1,FKS}, the connecting lines between the circles are between the
intersection points of the circles with a line parallel to the real axis through
the center of the circle. This had advantages in the treatment of real curves,
but with the obvious effect that the connecting lines were not of minimal length
as they are now. In addition they could enter the interior of the circles and
thus come closer to the critical points than the distance $R$. In the present
approach, both these issues are addressed.


The numbering of the problem points is as indicated by the order in which the
points are identified as discussed in the previous section; only the point with
the base point is swapped with the point no. 1.  To enforce relation
(\ref{gamma}), as in \cite{FKS} the relative position of the $\gamma_{i}$,
$i=1,\ldots,N_{c}$ has to be determined. To this end the \emph{endpoints} of the
spanning tree, i.e., the points $b_{i}$ with $i>1$ just connected to one other
point, are determined, in Fig.~\ref{fig:monpaths} the points 3, 6, 5, 2. In
addition the \emph{nodes} are identified, i.e., the points connected (possibly
via intermediate points) to several endpoints, in Fig.~\ref{fig:monpaths} the
points 4 and 1. Then starting from the endpoint all branches are traced to the
next node. At a node, the branches are ordered according to the angles of the
connecting lines.  This procedure is repeated until the base point is reached
where a final ordering according to the angles of the connecting lines with
respect to the arriving line (the line to the next point on the lower level of
the spanning tree) is performed. For Fig.~\ref{fig:monpaths}, this leads to the
sequence 3, 6, 5, 2, 4, 1. In contrast to \cite{FKS}, we do not change the
contours $\gamma_{i}$ ($i=1,\ldots,N_{c}$), but permute the problem points to
enforce (\ref{gamma}). This leads to the order of the branch points given at the
end of Sect.~\ref{2}.

\section{Computation of monodromies and periods}%
\label{5}

In this section we identify the monodromy group of the covering of the Riemann
sphere regarded as the projective space $\mathbb{C}P^1$ as given by
(\ref{Fdef}).  An algebraic curve (\ref{fdef}) defines an $N$-sheeted covering
of the Riemann sphere. This covering can be characterized by the following data:
branch points and permutations, which are called the monodromies of the
covering. To compute the monodromies, we lift the basis of
$\pi_1(\mathbb{C}P^1\setminus{\{b_i\}_{i=1}^{N_c}})$, the contours $\gamma_{i}$,
$i=1,\ldots,N_{c}$ constructed in the previous section, to the covering.

\subsection{Monodromy group}

Since by construction no branch point or singularity nor points, where~$y$
diverges, lie on the $\gamma_{i}$, there are always $N$ distinct finite roots
$y_{n}$ of $f(x,y)=0$ for a given $x\in \gamma_{i}$, $i=1,\ldots,N_{c}$. The
procedure is similar to the one in the Maple \name{algcurves} package and in
\cite{alg1}: At the base point $a$ we enumerate the sheets and obtain a vector
$\mathrm{y}(a)=(y_{1}(a),\ldots,y_{N}(a))=:(A_{1},\ldots,A_{N})$ by solving
$f(a,y)=0$. If we start at a point $A_{k}$, $k=1,\ldots,N$ on the covering, and
consider the analytic continuation of the vector of roots $\mathrm{y}$ along one
of the contours $\gamma_{i}$, we will obtain a permutation of the components of
the vector $\mathrm{y}$ back at the base point,
\begin{equation}
    \sigma_{i}\mathrm{y}:=(y_{\sigma_{i}(1)}(a),\ldots,y_{\sigma_{i}(N)}(a))
    \label{perm}
\end{equation}
The permutation $\sigma_{\infty}$ associated with $x=\infty$ can be computed in
the same way along a contour $\gamma_{\infty}$ with negative orientation
surrounding all finite branch points, i.e., a contour for which we have relation
(\ref{gamma}). Alternatively, it follows from the permutations obtained for the
finite discriminant points via the relation
\begin{equation}
    \sigma_{\infty}\circ \sigma_{Nc}\circ\ldots\circ\sigma_{1}=1
    \label{sigma}\;.
\end{equation} 
The group generated by the $\sigma_{i}$ is called the monodromy group
of the covering.

\subsection{Collocation methods}

As in \cite{alg1}, we numerically construct the analytic continuation of the
vector $\mathrm{y}$ along the lifted contours $\gamma_{i}$ and compute at the
same time the integrals of the holomorphic differentials along the $\gamma_{i}$,
$i=1,..,N_{c}$.  Thus, the collocation points on the contours are chosen in
accordance with the numerical integration scheme which is here the
\emph{Clenshaw-Curtis method}.  The theoretical basis of this method is an
expansion of the integrand in Chebychev polynomials by a \emph{collocation
  method} on the \emph{Chebychev points} $x_{i}=\cos(i\pi/N_{l})$,
$i=1,\ldots,N_{l}$.  This means we approximate the function $\mathcal{F}(x)$ for
$|x|\leq 1$ by an interpolating polynomial written as a sum of Chebychev
polynomials $T_{n}(x)$: $\mathcal{F}(x)\sim \sum_{k=0}^{N_{l}}a_{k}T_{k}(x)$,
where the \emph{spectral coefficients} $a_{k}$ follow from imposing this
approximation as an exact condition at the Chebychev points, i.e.,
$\mathcal{F}(x_{l})=\sum_{k=0}^{N_{l}}a_{k}T_{k}(x_{l})$, $l=0,\ldots,N_{l}$
($N_{l}$ is also called the number of \emph{modes} in the computation).
Consequently, we obtain
\begin{equation}
    \int_{-1}^{1}\mathcal{F}(x)dx\sim 
    \sum_{k=0}^{N_{l}}a_{k}\int_{-1}^{1}T_{k}(x)\D x
    \label{leg}\;.
\end{equation}
An expansion of a function with respect to a system of globally smooth functions
on their domain is called a \textit{(pseudo-)spectral method}.  The computation
of the spectral coefficients $a_{k}$ by inverting the matrix $T_{k}(x_{l})$ and
the integral of the Chebychev polynomials in (\ref{leg}) can be combined in the
so-called Clenshaw-Curtis weights $w_{k}$, with which (\ref{leg}) can be written
in the form
\begin{equation}
    \int_{-1}^{1}\mathcal{F}(x)\D x\sim 
    \sum_{k=0}^{N_{l}}\mathcal{F}(x_{k})w_{k}
    \label{legw}\;.
\end{equation}
Thus, for given function values $\mathcal{F}(x_{k})$ at the Chebychev points
$x_{k}$ and weights $w_{k}$, the numerical approximation of the integral is just
the computation of a scalar product. The weights can be conveniently determined
in Matlab via Trefethen's code~\cite{trefethen1,trefethenweb}. They have to be
computed only once and are then stored for later use in the numerical
integrations.  This is one of the reasons why the same radius of the circles and
thus the same mappings to the interval $[-1,1]$ and the same number $N_{l}$ of
Chebychev points are used for all circles and lines (on the connecting lines,
typically one fourth of the points on the circles are used, but this can be
freely chosen).

In contrast to \cite{alg1}, where only half-circles were considered, the
computation of the integrals along the contours $\gamma_{i}$, $i=1,\ldots,N_{c}$
also includes integrations over segments of a circle. The integration along
these segments is done by expanding the integrands in terms of Chebychev
polynomials as in \eqref{leg}. Then the well known identity
\begin{equation}
    \frac{T_{n+1}'(x)}{n+1}-\frac{T_{n-1}'(x)}{n-1}=2T_{n}(x),\quad n>0,
    \label{chebint}
\end{equation}
is used to determine the coefficients of the expansion of the 
anti-derivative of  $\mathcal{F}(x)$ in terms of  Chebychev 
polynomials. This gives the value of the integral over the segment.
Note that the spectral coefficients $a_{k}$ can be computed with a \emph{fast 
cosine transformation} (see for instance \cite{trefethen1}) which is 
the reason why we use Chebychev polynomials here instead of Legendre 
polynomials in \cite{alg1} for which no fast algorithm is known.
It would be of course possible to map all segments on the circle to 
the interval $[-1,1]$ and use Clenshaw-Curtis there, but this is 
not done to allow for a better vectorization of the code.

It is known that the difference between an analytic function $\mathcal{F}$ and
its spectral approximation decreases exponentially with $N_{l}$. Here we have to
integrate functions of $x$ and $y$ on a set of contours where the functions are
analytic, which guarantees an optimal efficiency of the method provided the
radius $R$ of the circles is not too small. Thus, we can reach machine precision
typically with $N_{l}\leq 2^{8}$. The contours consist of the circle segments
discussed above, and of lines and full circles each of which is mapped to the
interval $[-1,1]$, where we use Clenshaw-Curtis integration. To reach machine
precision, it is obviously necessary to know the integrand with this precision.
Therefore we solve the algebraic equation $f(x,y)=0$ on each Chebychev point.
Since the $\gamma_{i}$, $i=1,\ldots,N_{c}$, by construction do not come close to
branch points or singularities, no multiple roots will occur. Thus, we can use
the Matlab command \texttt{roots} efficiently to determine $\mathrm{y}$. The
analytic continuation is obtained by sorting the newly computed vector
components according to minimal difference with the components at the previous
collocation point.  Carrying out this procedure starting from a base point
$A_{k}$ along some contour $\gamma_{i}$, $i=1,\ldots,N_{c}$, one obtains the
permutation by comparing the analytic continuation of $\mathrm{y}$ along
$\gamma_{i}$ and $\mathrm{y}$.

\subsection{Periods of the holomorphic 1-forms}

Since $\mathrm{y}$ is then known at the Chebychev points, the same holds for the
differentials there. It is well known (see \cite{BK86,Noe83}) that the
holomorphic differentials on the Riemann surface associated to an algebraic
curve (\ref{fdef}) can be written in the form
\begin{equation}
    \omega_{k}=\frac{P_{k}(x,y)}{f_{y}(x,y)}\D x\;,
    \label{holdiff}
\end{equation}
where the adjoint polynomials
$P_{k}(x,y)=\sum_{i+j\leq d-3}^{}c^{(k)}_{ij}x^{i}y^{j}$ are of degree at most
$d-3$ in $x$ and $y$. If the curve has no singular points, there are no further
conditions on the $P_{k}$ and, consequently, there are $\tilde{g}=(d-1)(d-2)/2$
linearly independent polynomials $P_{k}$. Since, as is well known, the dimension
of the space of holomorphic 1-forms is equal to the genus $g$ of the Riemann
surface, the genus is $(d-1)(d-2)/2$ in this case.  If there are singular
points, the set of which is denoted by $S$, there is a number $\delta_{P}$ ---
called the \textit{delta invariant} --- of further conditions on the $P_{k}$ at
a point $P\in S$ as a consequence of the holomorphicity of the differentials
also at these points.  These will be discussed in Sect.~\ref{3a}.

Since we are interested in an efficient computation of the Riemann matrix, we do
not separate the monodromy computation from the integration of the holomorphic
differentials along these contours, but do both in one go. This also explains
why we use a much higher number of collocation points than needed for the
determination of the monodromies alone, the choice of the number $N_{l}$ is in
accordance with the resolution required for the determination of the periods.
As will be seen below, not all of these integrals are linearly independent. Our
integration procedure will thus provide more integrals than actually needed. But
the efficient vectorization algorithms in Matlab ensure that the computation of
the additional integrals will not be time consuming. In addition, the
possibility to check the validity of identities between the computed integrals
will provide strong tests of the numerical results since the integrals are
computed independently.

The monodromies $\sigma_{i}$ are stored in an $N\times N_{c}$-array. We then
check which of the discriminant points are actually branch points, i.e., have
non-trivial monodromy.  The monodromy at infinity is computed via (\ref{sigma})
from the monodromies at the finite branch points.  It can be computed via the
contour $\gamma_{\infty}$ as in Maple to provide an additional test, but this is
only done here if the Puiseux expansion at the infinite points is required or if
the Abel map for points in the vicinity of the infinite points is needed.  This
is done for reasons of numerical efficiency since more resolution is necessary
in general on the big contour surrounding all finite critical points, and since
all other information is already easily accessible at this stage.  The base
point used by the code is stored in the variable \texttt{base}, the vector
$\mathrm{y}(a)$ indicating the labeling of the sheets in the variable
\texttt{ybase}, the branch points in \texttt{bpoints}, and the monodromies in
the variable \texttt{Mon}.

\subsection{Example}

For the curve~(\ref{cubic}) the code produces the base point
\begin{verbatim}
base =
       -0.6778
ybase =
       -0.8374
        0.7299
        0.1075,
\end{verbatim}
the branch points 
\begin{verbatim}
bpoints =
  -0.3197 - 0.9839i
   0.8370 - 0.6081i
   0.8370 + 0.6081i
  -0.3197 + 0.9839i
  -0.0000 + 0.0000i
  -1.0346 + 0.0000i
      Inf + 0.0000i
\end{verbatim}
and the monodromies
\begin{verbatim}
Mon =
     3     1     3     1     2     1     3
     2     3     2     3     1     3     1
     1     2     1     2     3     2     2.
\end{verbatim}
\begin{equation}    
    \label{mono}
\end{equation}
This example shows that going around the first branch point one ends up in the
third sheet when starting in the first and vice versa, whereas sheet two is not
affected.

\begin{remark}
  The above procedure to compute monodromies and periods is insensitive to the
  accuracy with which the branch points are computed as long as the error in the
  location of the branch points is much smaller than the radius of the circles
  around the points. For numerical accuracy of the periods, it is just important
  that the branch points are not close to the contours.
\end{remark}

\section{Puiseux Expansions}
\label{3}
Whereas algebraic curves can have singularities as discussed in Sect.~\ref{2},
Riemann surfaces are smooth manifolds. In order to define a Riemann surface via
algebraic curves with singularities, a desingularization procedure has to be
implemented. As in the Maple \name{algcurves} package and in \cite{alg1}, we
construct an atlas of local coordinates for the Riemann surface corresponding to
the algebraic curve via series $y(x)$ with rational exponents in the vicinity of
the singular point. These \emph{Puiseux expansions} are calculated at least up
to the order necessary to identify all sheets of the Riemann surface near the
singularities. They are used as local coordinates in the vicinity of these
points which will provide part of an atlas for the description of the Riemann
surface as a smooth manifold. In \cite{RSbookdp,alg1}, Puiseux expansions were
constructed via the Newton polygon. To avoid as much as possible cancellation
errors, we use here an approach based on complex contour integrals. Duval
\cite{duval} and Poteaux \cite{poteauxj} gave efficient Puiseux expansion over
certain algebraic fields as rational numbers which are, however, not helpful in
the context of a purely numerical approach.

\subsection{Puiseux expansion at finite points}

A Puiseux expansion near a point $(x_{s},y_{s})$ on the curve in general
position is just a Taylor expansion of $y$ in terms of $x$.  In the vicinity of
a branch point or a singularity with non-trivial monodromy as determined in the
previous section, the corresponding series is entire only in the local parameter
$x=x_{s}+t^{r}$, where $r$ is the length of the monodromy cycle
$(k_{1}k_{2}\ldots k_{r})$, $k_{i}\in \mathbb{N}$, at the point $(x_{s},y_{s})$,
i.e., the number of sheets connected at this point.  We write the Puiseux
expansions in the form
\begin{equation}
    x=x_{s}+t^{r}\;, \quad y=y_{s}+\sum_{n=1}^{\infty}\alpha_{n}t^{n}\;,
    \label{puidef}
\end{equation}
where $r\in \mathbb{N}$, and where $(\alpha_{i})_{i=1,2,\ldots}\in
\mathbb{C}$.
Note that due to the definition of the local parameter $t$, there are in fact
$r$ expansions of the form (\ref{puidef}) which differ only by a change
$t\mapsto \epsilon t$, where $\epsilon$ is a root of unity of order $r$.  To
identify all sheets in the vicinity of the singular point $x_{s}$, $N$
inequivalent expansions of the form (\ref{puidef}) are needed.  The minimal
number of terms in the series to achieve this is called the \emph{singular part}
of a Puiseux expansion.

In the Maple \name{algcurves} package and in \cite{alg1}, the Puiseux expansion
(\ref{puidef}) is computed via the Newton polygon. To this end, the curve
(\ref{fdef}) is written in terms of the variables $\tilde{x}=x-x_{s}$,
$\tilde{y}=y-y_{s}$ as $\tilde{f}(\tilde{x},\tilde{y})=0$, i.e., as a curve for
which the expansion is studied at the point $(0,0)$. A disadvantage of this
approach for a purely numerical scheme working with inexact coefficients of the
polynomial (\ref{fdef}) and thus with inexact representations of $(x_{s},y_{s})$
is that these inaccuracies are amplified in the expansion of terms of the form
$(\tilde{x}+x_{s})^{i}$ and $(\tilde{y}+y_{s})^{j}$ because of possibly large
binomial coefficients. This problem also exists in the approach with contour
integrals, since $f(x,y)$ has to be evaluated on the circles around the point
$x_{s}$, but there the algebraic equation (\ref{fdef}) is solved for $y$ for
given $x$ with machine precision. A worse problem for the numerical approach via
the Newton polygon are cancellation errors if the singularity is of high order
since the first terms in the expansion (\ref{puidef}) will cancel. In
\cite{alg1} we discussed the example of the curve
\begin{equation}
    f(x,y)=((y^3+x^2)^2+x^3y^2)^2+x^7y^3=0
    \label{highsing}
\end{equation}
having a singularity at $(0,0)$, where the singular part of the Puiseux
expansion consists of 3 terms. In this case the rounding errors in the Puiseux
coefficients introduced errors of the order of $10^{-5}$ in the conditions on
the adjoint polynomials and thus in the holomorphic differentials to be
constructed in the next section.

To avoid as much as possible these numerical problems, we use the well known
Cauchy formula to compute the coefficients of a Taylor expansion of a
holomorphic function. The procedure outlined below also works for general points
$(x_{s},y_{s})$, where first $y$ has to be determined as discussed in the
previous section on a small circle centered at $x_{s}$ in the base. For the
branch points and singularities to be discussed in the following, this has been
done already during the computation of the monodromy group which makes the
approach particularly efficient.  For the point $(x_{s},y_{s})$ it is determined
to which monodromy cycle $(k_{1}k_{2}\ldots k_{r})$, $k_{i}\in \mathbb{N}$ it
belongs. Denoting the lifts of the circle of radius $R$ around $x_{s}$ in the
base to the corresponding sheets in the covering by
$\Gamma_{k_{1}},\ldots, \Gamma_{k_{r}}$ respectively, the contour
$\Gamma=\Gamma_{k_{1}}\circ\ldots\circ\Gamma_{k_{r}}$ is a closed contour on the
covering. Thus the Cauchy formula can be used in standard way to obtain the
coefficients $\alpha_{n}$ in (\ref{puidef}) via
\begin{equation}
    \alpha_{n}=\frac{1}{2\pi 
    \mathrm{i}}\int_{\Gamma}^{}y(t)\frac{\mathrm{d}t}{(1-t_{0}/t)^{n}t^{n+1}},\quad 
    n=1,2\ldots,
    \label{alphant0}
\end{equation}
if the expansion  is done at a point $t_{0}$ different from the center of the 
circle.

Parametrizing in (\ref{alphant0}) the circle in the base by
$x=x_{s}+R\exp(\mathrm{i}\phi)$, $\phi\in[0,2r\pi]$ (the index of the contour is $r$), we
get for $t_{0}=0$
\begin{equation}
    \alpha_{n}=\frac{1}{2\pi rR^{n/r}}\sum_{k\in(k_{1}\ldots k_{r})}
    \int_{0}^{2\pi}y(t)\exp(-\mathrm{i}\phi 
    n/r-2\pi\mathrm{i}n(k-1)/r)\,\mathrm{d}\phi, \quad n=1,2,\ldots
    \label{alphan2}.
\end{equation}
Since the integrands are known on Chebychev collocation points for $\phi$, the
Clenshaw-Curtis algorithm (\ref{legw}) can be applied as before with negligible
computational cost. The coefficients $\alpha_{n}$, $n=1,\ldots,N_{p}$ are
computed in a vectorized way for all desired values of $n$ at the same time. The
value of $N_{p}$ can be freely chosen (to allow for highly degenerate cases, the
default value is $N_{p}=100$).  Values of the coefficients of the Puiseux
expansion at some point $x_{s}$ are given for all cycles of the monodromy.

For the example (\ref{cubic}), the point $(0,0)$ is, as already mentioned, a
singular point, and the Puiseux expansions in the vicinity of the point are
$x=t$, $y=t^{4}/2-t^{9}/16+\ldots$ and $x=t^{2}$,
$y=-\sqrt{2}\mathrm{i}t^{3}-t^{8}/4+\ldots$.  The code produces the following
result,
\begin{verbatim}
    PuiExp{1}(:,1:11) = 

Columns 1 through 4

   2.0000 + 0.0000i  -0.0000 + 0.0000i   0.0000 + 0.0000i   0.0000 + 0.0000i
   1.0000 + 0.0000i   0.0000 + 0.0000i  -0.0000 - 0.0000i  -0.0000 + 0.0000i

  Columns 5 through 8

   0.0000 - 1.4142i   0.0000 + 0.0000i   0.0000 + 0.0000i  -0.0000 + 0.0000i
   0.0000 - 0.0000i   0.5000 + 0.0000i   0.0000 + 0.0000i  -0.0000 - 0.0000i

  Columns 9 through 11

   0.0000 + 0.0000i  -0.2500 - 0.0000i  -0.0000 - 0.0000i
  -0.0000 + 0.0000i   0.0000 + 0.0000i  -0.0625 + 0.0000i
  \end{verbatim}
where the first column of the output gives the length $r$ of the monodromy cycle
(recall that as in the Maple \name{algcurves} package there are always $r$
expansions differing from the ones shown by multiplication of $t$ with roots of
unity of order $r$). Note that the first given coefficients in the second column
correspond to $n=0$.

\begin{figure}[htb]
    \centering 
    \includegraphics[width=10cm]{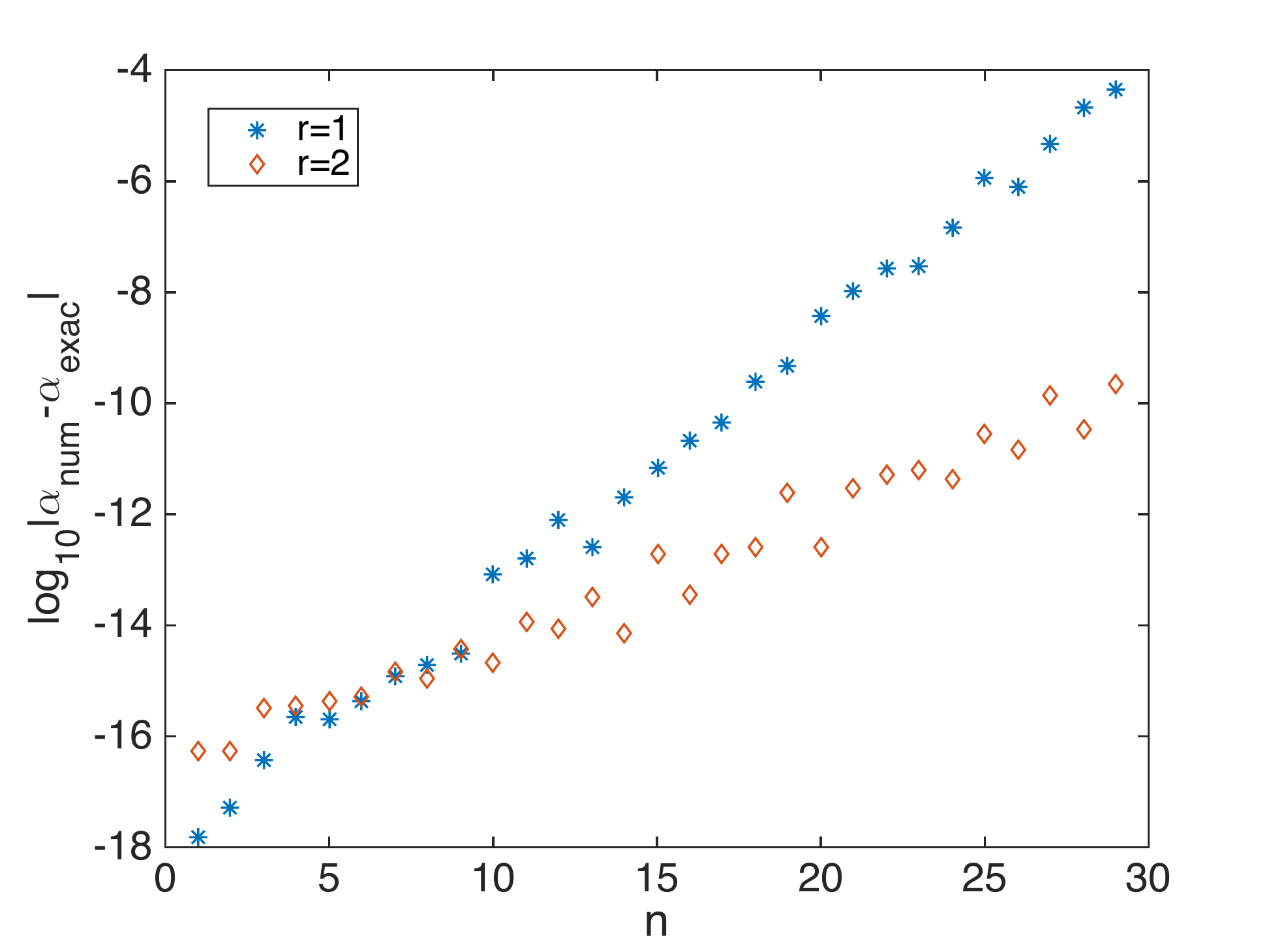}
    \caption{Difference of the numerically computed and the exact Puiseux 
    coefficients for the curve (\ref{cubic}) in a logarithmic plot.}
    \label{fig:puicoeff}
\end{figure}

If there is sufficient resolution in terms of Chebychev points to compute the
integrals in (\ref{alphan2}), this will be done with the Clenshaw-Curtis
algorithm to machine precision. But since the radius $R$ is in general smaller
than 1, the numerical errors are growing with $n$ as $R^{-n/r}$. This can be
seen for the example (\ref{cubic}), where $R=0.3567$, at the singularity $(0,0)$
in Fig.~\ref{fig:puicoeff}, where the difference between the exact and the
numerical Puiseux coefficients is shown in a logarithmic plot. Since $1/R^{30}$
is of the order $10^{13}$, the error in $\alpha_{30}$ is for $r=1$ of the order
of $10^{-4}$, whereas for $r=2$ it is still of the order of $10^{-10}$. Thus the
size of $R$, which can be at most half the minimal distance to the closest
branch point, delimits the accessible accuracy of the Puiseux coefficients in
this finite precision approach.

\subsection{Puiseux expansion at infinite points}

To obtain the Puiseux expansion at points on the curve where $x$ is infinite,
one could use the information along the contours around the finite critical
points as it was done for the computation of the monodromy at infinity in the
previous section. However, this would imply to compute integrals of the form
(\ref{alphant0}) also along the connecting lines between the circles and to
trace roots of $x-x_{s}=t^{r}$ there. Since the Matlab root is always branched
along the negative real axis, this would require an analytic continuation of the
roots along these lines as in the previous subsection. This can be done, but it
is cumbersome which is why we use other approaches here.

If the curve has singularities at infinity as determined in section \ref{2}, we
consider Puiseux expansions at infinity in the homogeneous coordinates of
Sect.~\ref{2} and use the same approach as for the finite singular points above
at the point $(X_{s},Y_{s},0)$. We determine the critical points of the curve
$F(X,Y,Z)=0$ (\ref{Fdef}) (either $X=1$ or $Y=1$) to establish the minimal
distance between discriminant points for this curve to identify a possible
radius $R$ for the circle around $(X_{s},Y_{s},0)$. Then equation (\ref{Fdef})
is solved for $Z$ in dependence of $X$ (if $Y=1$) or for $Y$ in dependence of
$Z$ (if $X=1$) as discussed in the previous section. This approach has the
additional benefit to provide an independent check of the monodromy computation
along the contours around the finite critical points. The Puiseux coefficients
are then determined via formula (\ref{alphan2}).

As an example we consider the curve (\ref{cubic}) which has a singular point
$(0,1,0)$. The curve $F(X,Z)=Z^{4}+2X^{3}Z^{3}-X^{7}=0$ has a singularity at
$(0,0)$, where all four sheets are connected. The code gives the Puiseux
expansion (again there are four expansions of this form differing by
multiplication of $t$ with a fourth root of unity)
\begin{verbatim}
    PuiExp{2}(1:11) = 

Columns 1 through 4

   4.0000 + 0.0000i   0.0000 - 0.0000i  -0.0000 + 0.0000i  -0.0000 + 0.0000i

  Columns 5 through 8

  -0.0000 - 0.0000i  -0.0000 + 0.0000i  -0.0000 + 0.0000i  -0.0000 + 0.0000i

  Columns 9 through 11

  -1.0000 - 0.0000i   0.0000 + 0.0000i   0.0000 + 0.0000i.
\end{verbatim}
The first entry gives again the value of the length of the monodromy 
cycle, $r=4$. This shows that the expansion is of the form 
$Z=\alpha_{7}t^{7}$ which implies in leading order for (\ref{cubic}) that 
$x=1/(\alpha_{7}t^{3})$ and $y=1/(\alpha_{7}t^{7})$. Thus we recover 
that the 3 sheets are connected at infinity, and that $y\propto 
x^{-7/3}$ there.

As an option, the code offers the possibility to compute the Puiseux expansion
at infinity directly in the coordinates $x$ and $y$. To this end, a circle of
radius $R_{\infty}$ centered at the origin, surrounding all circles of radius
$R$ around the critical points and touching at least one of these circles is
constructed. At the point, where the two circles touch, the roots for $y$ have
already been determined. If the circle is around the branch point $b_{n}$, then
the points $b_{1},\ldots,b_{n-1}$ have already been surrounded before reaching
the touching point along the contours constructed in section \ref{4}. Thus the
computed monodromies for these contours have to be applied to the value of $y$
at the touching point. These values of the roots of $y$ are then analytically
continued along the circle in positive orientation as before. This is again done
between Chebychev collocation points on the circle. By construction the computed
monodromy must be equal to $\sigma_{\infty}^{-1}$ computed via
(\ref{sigma}). The code reports an error if this is not the case.

Given the values of $y$ on the Chebychev collocation points on the circle of
radius $R_{\infty}$, the Puiseux expansion can be computed in the same way as
for the finite critical points in the previous subsection. The only difference
is that negative powers in $1/x^{1/r}$ can appear in this case. Note that we
use here in general twice the resolution on the circle with radius $R_{\infty}$
than for the circles with radius $R$, simply because $R_{\infty}$ is by
construction bigger than $R$. For the example (\ref{cubic}), we find (the first
value corresponds to the power $t^{-7}$, the second to $t^{-6}$ and so on)
\begin{verbatim}

  Columns 1 through 4

  -0.5000 - 0.8660i  -0.0000 - 0.0000i  -0.0000 - 0.0000i  -0.0000 + 0.0000i

  Columns 5 through 8

  -0.0000 + 0.0000i   0.3333 - 0.5774i   0.0000 - 0.0000i  -0.0000 + 0.0000i
\end{verbatim}
This corresponds to $t^{-7}-2t^{-2}/3$ where 
$t=\exp(2\pi \mathrm{i}/3)/x^{1/3}$, i.e., the same result as obtained above 
by using the projective coordinates ($X,Y,Z$).

\section{Basis of the holomorphic differentials on the Riemann Surface}%
\label{3a}

In this section, we describe the construction of a basis of the holomorphic 1-forms
on the Riemann surface if the set $S$ of singular points is non-empty. To this
end, we will use the complex contour integral techniques of the previous
section.  As already discussed in Sect.~\ref{5}, the holomorphic differentials
are of the form (\ref{holdiff}) with
$P_{k}(x, y) = \sum_{i+j\leq d-3}^{} c^{(k)}_{ij}x^{j}y^{j} $. At the singular
points, we use a Puiseux expansion in the local parameter $t$ of
(\ref{alphant0}) near the singularity of the integrands (\ref{holdiff}). For the
integral to be holomorphic, there must be no negative powers of $t$ in this
expansion which leads to a number $\delta_{P}$, the \emph{delta invariant}, of
conditions at each singular point $P$.  The genus of the surface is given in
this case by
\begin{equation}
    g=\frac{1}{2}(d-1)(d-2)-\sum_{P\in S}^{}\delta_{P}
    \label{genus}\;.
\end{equation}


To determine the conditions on the differentials (\ref{holdiff}) at the
corrected singular point $(x_{s},y_{s})$, we determine the Puiseux expansion of
$f_{y}$ which has now (within numerical precision) a vanishing constant
term. Since we are interested in the highest power of $t$ appearing in the
denominator of (\ref{holdiff}), i.e., the smallest power $N_{f}$ of $t$
appearing in the Puiseux expansion of $f_{y}$, we consider the formula
(\ref{alphan2}) with $y$ replaced by $f_{y}$.

A differential of the form (\ref{holdiff}) is holomorphic at a singular point
$(x_{s},y_{s})$ if no negative powers of $t$ in the Puiseux expansion of the
integrand appear. To establish the conditions on the differentials to ensure
this, we multiply the differential by powers $t^{k}$, $k=0,\ldots,N_{f}-1$ and
compute the integral along the contour $\Gamma$ of (\ref{alphant0}). The residue
theorem implies that the terms in the Puiseux expansion of the differentials
proportional to $1/t$ lead to a non-vanishing integral.  Singular points at
infinity are treated in a completely analogous way as in the last section as
Puiseux expansions at infinity in the homogeneous coordinates $(X,Y,Z)$.

To implement the conditions on the adjoint polynomials at the singularities in
Matlab, we write the matrix $c^{(k)}_{ij}$ with $i+j\leq d-3$ in standard way as
a vector $\mathrm{c}$ of length $(d-1)(d-2)/2$.  The holomorphicity of
(\ref{holdiff}) at the singular points implies relations of the form
$H\mathrm{c}=0$ where $H$ is a
$((d-1)(d-2)/2)\times \sum_{P\in S}^{} \delta_{P}$ matrix.  Each condition on
$\mathrm{c}$ following from a non-vanishing residue in the above expansions
gives a line in $H$. The first such condition found is stored as the first row
of $H$.  For each subsequent condition found, it is checked that this new
condition is linearly independent of the already present ones in $H$. This is
done in the following way: if the matrix $H$ contains $M$ linearly independent
conditions on $\mathrm{c}$, the first $M$ rows will be non-trivial, and $H$ has
rank $M$. The new condition will be tentatively added as row $M+1$ in $H$, and
it will be checked if the resulting matrix has rank $M+1$. If not, the new
condition is linearly dependent on the $M$ conditions already stored in $H$, and
line $M+1$ will be suppressed. At the end of this procedure, there will be
$\sum_{P\in S}^{}\delta_{P}$ non-trivial lines in this matrix. The holomorphic
differentials correspond to the vectors $\mathrm{c}$ in the kernel of the matrix
$H$. They are determined with the Matlab command \name{null}, where
\name{null}(H) provides an orthonormal basis for the null space of $H$.  Notice
that for reasons of numerical accuracy we do not look for a rational basis
$\mathrm{c}$ of the kernel of $H$ even in cases where such a basis exists. The
polynomials $P_{k}$ are stored in the form of matrices $c^{(k)}_{nm}$ where in
Matlab convention $P_{k}(x,y)=\sum_{n,m=0}^{d-3}c^{(k)}_{nm}x^{d-3-n}y^{d-3-m}$,
and where the first row/column has $n=0/m=0$. The code gives the matrices
$c^{(k)}$ as $c\{1\},\ldots,c\{g\}$, where $g$ is the genus of the Riemann
surface.  For the curve (\ref{cubic}), where $d=7$, we get
\begin{verbatim}
c{1} =
         0     0     0     0     0
         0     0     0     0     1
         0     0     0     0     0
         0     0     0     0     0
         0     0     0     0     0
c{2} =
        Columns 1 through 4

   0.0000 + 0.0000i   0.0000 + 0.0000i   0.0000 + 0.0000i   0.0000 + 0.0000i
   0.0000 + 0.0000i   0.0000 + 0.0000i   0.0000 + 0.0000i   0.0000 + 0.0000i
   0.0000 + 0.0000i   0.0000 + 0.0000i   0.0000 + 0.0000i   0.0000 + 0.0000i
   0.0000 + 0.0000i  -0.0000 + 0.0000i  -0.0000 + 0.0000i   0.9938 - 0.1111i
  -0.0000 + 0.0000i   0.0000 + 0.0000i   0.0000 + 0.0000i   0.0000 + 0.0000i

  Column 5

  -0.0000 + 0.0000i
   0.0000 + 0.0000i
   0.0000 + 0.0000i
   0.0000 + 0.0000i
   0.0000 + 0.0000i
\end{verbatim}
i.e., the polynomials $P_{1}=xy$ and $P_{2}=x^{3}$.  Thus, the curve has genus
$2$ which implies it is \emph{hyperelliptic} since all surfaces of genus 2 are
(recall that hyperelliptic surfaces can be written in Weierstrass normal form
$\tilde{y}^{2}=\mbox{polynomial}(\tilde{x})$ after a birational transformation
$\tilde{x}(x,y)$, $\tilde{y}(x,y)$).

The integrals of the holomorphic differentials along $\gamma_{i}$ are obtained
as discussed above, and the results are stored in an
$N\times N_{c}\times \tilde{g}$ array.  The sum of integrals of a holomorphic
differential over all contours with the same projection into the $x$-sphere must
vanish.  In practice this sum will not vanish because of numerical errors and
thus gives an indication on the quality of the numerics.  The code issues a
warning if this sum is greater than the prescribed rounding tolerance
$\mathtt{Tol}$.

\section{Homology of a Riemann Surface}%
\label{6}

The monodromies computed in Sect.~\ref{5} provide the necessary information to
determine a basis for the homology on a Riemann surface. We use as in the Maple
\name{algcurves} package and in \cite{alg1} the algorithm by Tretkoff and
Tretkoff \cite{tret} to construct such a basis. We outline only the main steps
here, for details the reader is referred to \cite{tret} and \cite{Bob}.

The first step in the construction of the homology is the identification of the
points on the covering belonging to more than one sheet, i.e., the points, where
the\emph{ branching number} (the number of sheets to which a point belongs minus
1) is different from zero.  To this end one has to identify the cycles within
the permutations in the monodromies computed in Sect.~\ref{5}. This is simply
done by determining for each discriminant point with non-trivial monodromy,
i.e., each branch point, the sheets which are permuted whilst encircling this
point. The permuted sheets form the cycles within the permutation. They are
identified as follows: the monodromies are given as permutations of the vector
$(1,2,\ldots,N)$. For each permutation vector the code identifies the components
which are not in the order $(1,2,\ldots,N)$. For the first such component, it
goes to the sheet indicated by this component, then to the next indicated by the
component in the vectors there, until the starting point for the procedure is
reached again. This identifies the first cycle. For the first vector in the
example (\ref{mono}), the first permuted component corresponds to the first
sheet. The 3 there indicates that going around this point in the first sheet,
one ends up in the third. The third component of the permutation vector is a 1,
the sheet where this cycle $(1,3)$ started.

If not all permuted sheets appear in the first cycle identified in this way, the
procedure is repeated for the remaining permuted sheets until all permutation
cycles are determined.  Each such cycle corresponds to one of the $N_{B}$
ramification points on the covering and is labelled by $B_{i}$,
$i=1,\ldots, N_{B}$, where it is possible that several such points have the same
projection onto the complex plane.  From the number $n_{i}$ of elements in the
permutation cycle we obtain the branching number $\beta_{i}=n_{i}-1$.  The
Riemann-Hurwitz formula then allows the computation of the genus:
\begin{equation}
    g=\frac{1}{2}\sum_{i=1}^{N_{B}}\beta_{i}+1-N
    \label{RieHur}\;.
\end{equation}
Since the determination of the genus via monodromies is completely independent
from the genus computation via the dimension of the space of the holomorphic
1-forms, this provides a strong test for the code.  A failure in this test
results in an error which in general indicates that the branch points and
singularities or the holomorphic 1-forms were not correctly identified, the
latter typically for a lack of resolution.  For the curve (\ref{cubic}) we find
that all branch points on the covering connect exactly two sheets except for the
one above infinity where 3 sheets come together. Therefore, the genus is 2 in
accordance with the results for the holomorphic 1-forms.

The Tretkoff-Tretkoff algorithm constructs a spanning tree connecting the points
$A_{j}$ on the covering (the points projecting to the base point in the base)
starting from $\{A_{1}\}$ with the points $\{B_{i}\}$.  The algorithm allows to
identify non-trivial closed cycles on the Riemann surface and to compute
intersection numbers between them via a planar graph.  In total, one obtains $2g+N-1$ cycles. Since
the homology on a Riemann surface of genus $g$ has dimension $2g$, the cycles
cannot be all linearly independent.  A canonical basis of cycles
$a_{1},\ldots,a_{g}, b_{1},\ldots,b_{g}$ on a Riemann surface of genus $g$
satisfies the intersection conditions
\begin{equation}
    a_{i}\circ b_{j}=\delta_{ij},\quad a_{i}\circ a_{j}=0,\quad b_{i}\circ 
    b_{j}=0, \quad i,j=1,\ldots,g
    \label{canonical}.
\end{equation}
To find a canonical basis with the Tretkoff-Tretkoff algorithm, 
one has to compute the intersection matrix $\mathbf{K}$
for the obtained $2g+N-1$ cycles, which is straightforward to do from
the planar graph. It was shown in \cite{tret} that the resulting matrix has rank $2g$.
For the example of the curve
(\ref{cubic}) we obtain the intersection matrix:
\begin{verbatim}
K =
     0     1     0     0    -1    -1
    -1     0     0     1    -1    -1
     0     0     0     1    -1     0
     0    -1    -1     0     1     0
     1     1     1    -1     0     1
     1     1     0     0    -1     0.
\end{verbatim}
This matrix can be transformed to the canonical form
\begin{equation}
    \mathbf{\alpha} \mathbf{K} \mathbf{\alpha}^{T}=
    \begin{pmatrix}
        \mathbf{0}_{g} & \mathbf{I}_{g} & \mathbf{0}_{g,N-1}  \\
        -\mathbf{I}_{g} & \mathbf{0}_{g} &  \mathbf{0}_{g,N-1} \\
        \mathbf{0}_{N-1,g} & \mathbf{0}_{N-1,g} & \mathbf{0}_{N-1,N-1}
    \end{pmatrix}
    \label{inter}\;,
\end{equation}
where $\mathbf{\alpha}$ is a $(2g+N-1)\times(2g+N-1)$-matrix \ with
integer entries and $\det \mathbf{\alpha}=\pm1$, $\mathbf{0}_{g}$ is
the $g\times g$ zero matrix, $\mathbf{I}_{g}$ is the $g\times g$
identity matrix, and $\mathbf{0}_{i,j}$ the $i\times j$ zero
matrix. The canonical basis of the homology of the surface is given by
the cycles $a_{i}$ and $b_{i}$:
\begin{equation}
    a_{i}=\sum_{j=1}^{2g+N-1}\alpha_{ij}c_{j}\;, \quad 
    b_{i}=\sum_{j=1}^{2g+N-1}\alpha_{i+g,j}c_{j}\;, \quad i=1,\ldots,g
    \label{cycle}\;,
\end{equation}
where $c_{j}$ are the $2g+N-1$ closed contours obtained from the
planar graph.  The remaining cycles are homologous to zero,
\begin{equation}
    0=\sum_{j=1}^{2g+N-1}\alpha_{ij}c_{j}\;,\quad i=2g+1,\ldots,2g+N-1
    \label{cycle2}\;.
\end{equation}
For the curve (\ref{cubic}),
the code produces
\begin{verbatim}
acycle{1} =
         1     1     3     2     2     5     1
acycle{2} =
         1     1     3     6     2     5     1
bcycle{1} =
         1     1     3     4     2     5     1
bcycle{2} =
         3     1     3     2     2     5     1     3     3,
\end{verbatim}
where $\mathtt{acycle}\{i\}$ corresponds to $a_{i}$, and
$\mathtt{bcycle}\{i\}$ to $b_{i}$.  These numbers are to be read in
the following way: The numbers at odd positions in the cycle
correspond to the indices $j$ of $A_{j}$, $j=1,\ldots,N$, the numbers
at the even positions to the indices $i$ of the $B_{i}$,
$i=1,\ldots,N_{B}$.  In the above example, the cycle $b_{1}$ starts in
the first sheet, goes around $B_{1}$ to end in the third sheet, then
around $B_{4}$ to the second sheet, then around $B_{5}$ 
to come back to the first sheet. The code can give more
detailed information on the Tretkoff-Tretkoff tree and the cycles
$c_{k}$, $k=1,\ldots,2g+N-1$, as an option (it is stored in the
variable \texttt{cycle} in the code tretkoffalg.m).

Relations (\ref{cycle}) allow the computation of the periods of the 
holomorphic differentials from the integrals along the contours 
$\Gamma_{i}$ used for the monodromy computation. The cycles $c_{k}$, 
$k=1,\ldots,2g+N-1$, 
are equivalent to a sequence of contours $\Gamma_{i}$. Thus, using 
(\ref{cycle}) and the integrals of the holomorphic differentials 
along the contours $\Gamma_{i}$, we get  the $a$- and 
$b$-periods of the holomorphic differentials, the matrices $\mathcal{A}$ and 
$\mathcal{B}$, respectively. The Riemann matrix $\mathbb{B}$  is given by
\begin{equation}
    \mathbb{B}=\mathcal{A}^{-1}\mathcal{B}
    \label{riemat}.
\end{equation}
Since the Riemann matrix must be symmetric, the asymmetry of the
computed matrix is a strong test for the numerical accuracy.  A
warning is reported if the asymmetry is greater than the prescribed
tolerance. Similarly it is checked whether the periods along the
cycles (\ref{cycle2}) homologous to zero vanish with the same
accuracy.  For the curve (\ref{cubic}) the code finds the Riemann
matrix
\begin{verbatim}
RieMat =
   0.3090 + 0.9511i   0.5000 - 0.3633i
   0.5000 - 0.3633i  -0.3090 + 0.9511i.
\end{verbatim}
For a more compact representation we give only 4 digits here though 16
are available internally.  The $L^{\infty}$ norm  of $\mathbb{B}-\mathbb{B}^{T}$ 
and  of the
periods along the cycles (\ref{cycle2}) are  of the order of 
$10^{-15}$ with $2^{7}$ modes. We
will discuss the performance of the code in more detail in the next
section.

\section{Performance of the code}%
\label{7}
In this section we will illustrate the performance of the code for 
typical examples and discuss certain features of the chosen numerical 
approach.

We note that the code is constructed for high precision computation. 
The identification of the holomorphic differentials at singularities 
of the algebraic curve is done as the null space of conditions on the 
candidate differentials at the singularities. To be able to identify this null 
space (via the Matlab command \name{null}), these conditions obtained 
from complex contour integrals need to be known with sufficient 
precision. If this is not the case, the holomorphic differentials are 
not correctly found, and the code exits with an error due to a 
disagreement between the genus obtained via the monodromies and the 
dimension of the space of holomorphic differentials.
Due to the exponential decrease of the error in the
computation of the periods with the number $N_{l}$ of Chebychev 
polynomials, a general feature of spectral methods, this is in 
practice not a problem. But it explains why a precision of merely $10^{-3}$ 
cannot be reached with this code in the presence of singularities: 
the integrals have to computed with an error smaller than $10^{-6}$ 
in order that the holomorphic differentials are correctly identified.

\begin{figure}[htb]
    \centering 
    \includegraphics[width=10cm]{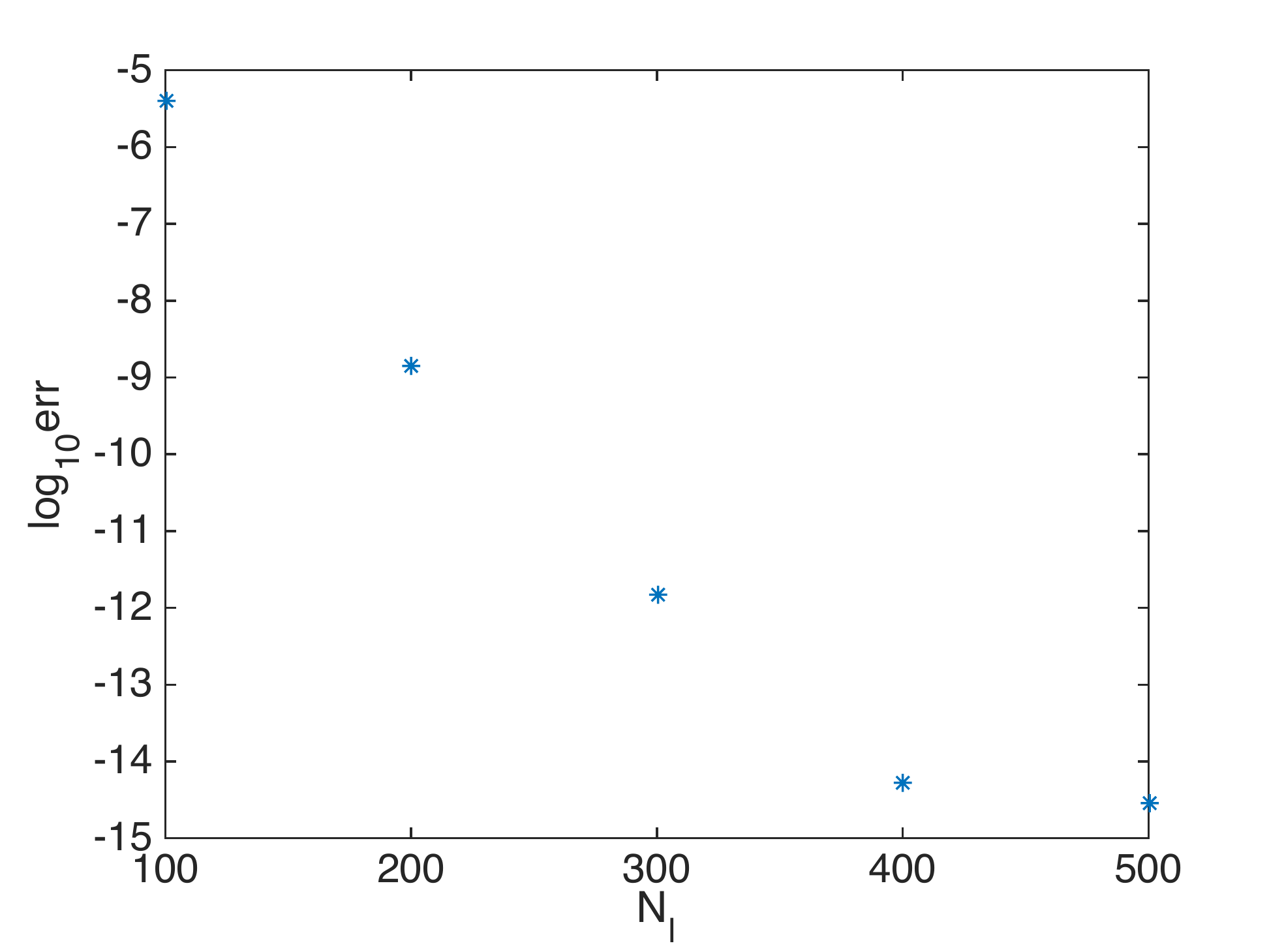}
    \caption{Numerical error \texttt{err} as defined in the text 
    for the curve  (\ref{nonic}).}
    \label{fig:error}
\end{figure}

The numerical
error we will study here in more detail is defined as the maximum of
the $L^{\infty}$ norm of the antisymmetric part of the numerically computed
Riemann matrix and the same norm of the right hand sides of
(\ref{cycle2}). The resulting variable is denoted by \texttt{err}.
For the curve (\ref{cubic}) we find that the code exits with an error 
with $N_{l}=2^{6}$ for the reasons outlined above. For 
$N_{l}=2^{7}$, the minimal error of order $10^{-15}$ is already 
reached. 
A more demanding test for the code is provided by the
curve
\begin{equation}
    f(x,y)=y^9+2x^2y^6+2x^4y^3+x^6+y^2=0,
    \label{nonic}
\end{equation}
which is a nine-sheeted genus 16 covering of the sphere with 42 finite branch 
points and two singular points $(0,0,1)$ and  (1,0,0). 
What makes this curve computationally challenging is the fact that the 
minimal distance between the branch points is just $0.018$. The 
dependence of the error on the number of Chebychev polynomials is 
shown in Fig.~\ref{fig:error}. It can be seen that machine precision 
is reached with 400 polynomials. 
The plot is typical for spectral methods: one
can see the exponential decrease of the error (an essentially linear
decrease in a logarithmic plot) and the saturation of the error once
machine precision is reached, here at roughly $10^{-14}$.

As already mentioned, the code \cite{alg1} had difficulties with 
curves having singularities where the
singular part of the Puiseux expansion consists of many terms due to
rounding problems occur. The example (\ref{highsing}) could only be 
computed with an error of the order of $10^{-5}$ because of this.  
This curve of genus 1 has a singularity at $(0,0)$ with a delta 
invariant 43, i.e., a singularity of high order with a singular part 
of the Puiseux expansion consisting of 3 terms. With $N_{l}=2^{7}$ 
Chebychev polynomials, the code exits with an error since the 
holomorphic differentials are not correctly identified at the 
singularity as discussed above. With $N_{l}=2^{8}$, an error of 
smaller than $10^{-14}$ is reached, a precision inaccessible with the 
code \cite{alg1}.

It is interesting to know for which operations the computation time in
Matlab is used.  Matlab timings have to be taken with a grain of salt 
since they depend on whether used functions are precompiled or not, 
but they are interesting from a practical point of view to get an 
idea how long a computation takes and what is the most time consuming 
task. We used the
vectorization algorithms in Matlab as much as possible to obtain an
efficient code.  For the Riemann matrix of the curve (\ref{cubic})
computed with 128 polynomials we find that roughly one third of the 
time  is used
for the analytic continuation of $\mathrm{y}$ along the contours and the
computation of the integrals (this includes solving the algebraic 
equation and sorting the roots to analytically continue the sheets, 
which both takes essentially the 
same time). The various Puiseux expansions needed in the determination 
of the holomorphic differentials also take roughly one third of the 
computing time (this is partly due to the fact that we work with a 
high number of Puiseux coefficients to allow for highly degenerate 
cases). The Clenshaw-Curtis integration,
which is just a matrix multiplication in this implementation, only
takes negligible computation time.  
The precise distribution of
computing time to the different numerical tasks necessary to obtain
the Riemann matrix for an algebraic curve depends of course on the
studied example. Note that the main advantage of the code presented 
here compared to the code \cite{alg1} is not speed, but efficiency. 
This means the former can treat cases with high precision where the 
latter fails or can only reach low accuracy. 

\section{Theta functions}%
\label{9}
Theta functions are the building blocks of  meromorphic functions
on Riemann surfaces. We define the theta
    function with characteristic     $\left [\mathrm{p},\mathrm{q}\right]$ 
    as an infinite series,
    \begin{equation}\label{theta}
    \Theta_{\mathrm{p}\mathrm{q}}(\mathrm{z},\mathbb{B})=
    \sum\limits_{\mathrm{N}\in\mathbb{Z}^g}\exp\left\{
    \I\pi\left\langle\mathbb{B}\left(\mathrm{N}+\mathrm{p}\right),
    \mathrm{N}+\mathrm{p}
    \right\rangle+2\pi \I
    \left\langle \mathrm{z}+\mathrm{q},\mathrm{N}+\mathrm{p}
    \right\rangle\right\}
    \;,
    \end{equation}
    with $\mathrm{z}\in\mathbb{C}^g$ and $\mathrm{p}$, $\mathrm{q}\in{
    \mathbb{R}}^g$, where 
    $\left
    \langle\cdot,\cdot\right\rangle$ denotes the Euclidean scalar product
    $\left\langle \mathrm{N},\mathrm{z}\right\rangle=\sum_{i=1}^gN_iz_i$.
The properties of the Riemann matrix ensure that the series converges
absolutely and that the theta function is an entire function on
$\mathbb{C}^{g}$. A characteristic is called \emph{singular} if the
corresponding theta function vanishes identically. 
 Of special importance are half-integer 
characteristics with $2\mathrm{p},2\mathrm{q}\in \mathbb{Z}^{g}$.  A 
half-integer characteristic is called \emph{even} if $4\langle 
\mathrm{p},\mathrm{q}\rangle=0\mbox{ mod } 2$ and \emph{odd} 
otherwise.
Theta functions with odd (even) characteristic are odd
(even) functions of the argument $\mathrm{z}$.  The theta function with
characteristic is related to the Riemann theta function $\Theta$, the
theta function with zero characteristic $\Theta:= \Theta_{\mathrm{00}}$,
via
\begin{equation}
    \Theta_{\mathrm{pq}}(\mathrm{z},\mathbb{B})=\Theta(\mathrm{z}
    +\mathbb{B}\mathrm{p} + \mathrm{q})\exp\left\{\I\pi
    \left\langle\mathbb{B}\mathrm{p},\mathrm{p}\right\rangle+
    2\pi \I\left\langle\mathrm{p},\mathrm{z} + \mathrm{q}\right\rangle
    \right\}\;.
    \label{thchar}
\end{equation}
The theta function has the periodicity properties 
\begin{equation}
    \Theta_{\mathrm{p}\mathrm{q}}(\mathrm{z}+\mathrm{e}_{j}) = 
    \mathrm{e}^{2\pi \mathrm{i}p_{j}}
    \Theta_{\mathrm{p}\mathrm{q}}(\mathrm{z})\;,
    \quad 
    \Theta_{\mathrm{p}\mathrm{q}}(\mathrm{z}+\mathbb{B}
    \mathrm{e}_{j})=
    \mathrm{e}^{-2\pi \I (z_{j}+q_{j}) - \I\pi B_{jj}}
    \Theta_{\mathrm{p}\mathrm{q}}(\mathrm{z})\;
    \label{eq:periodicity},
\end{equation}
where $\mathrm{e}_{j}$ is a vector in $\mathbb{R}^{g}$ consisting of
zeros except for a 1 in jth position.  In the computation of the
theta function we will always use the periodicity properties
(\ref{eq:periodicity}). This allows us to write an arbitrary vector
$\mathrm{z}\in \mathbb{C}^{g}$ in the form $\mathrm{z}=\hat{\mathrm{z}}+\mathrm{N}
+ \mathbb{B}\mathrm{M}$ with $\mathrm{N},\mathrm{M} \in \mathbb{Z}^{g}$, where
$\hat{\mathrm{z}}=\mathbb{B}\mathrm{p}+\mathrm{q}$ with $|p_{i}|\leq 
1/2$, $|q_{i}|\leq 1/2$,
and to compute $\Theta(\hat{\mathrm{z}},\mathbb{B})$ instead of
$\Theta(\mathrm{z},\mathbb{B})$ (they are identical up to an exponential
factor). The series (\ref{thchar}) will be approximated by a sum.  For details 
of the computation of the theta function, the reader is 
referred to \cite{cam} or \cite{alg1}.

\section{Abel map}%
\label{9a}
The Abel map $\omega:P\mapsto \int_{P_{0}}^{P}\mathrm{d}\omega$ 
is a bijective map from the Riemann surface $\mathcal{R}$ into the \emph{Jacobian} 
$Jac(\mathcal{R}):=\mathbb{C}^{g}/\Lambda$ where $\Lambda$ is the 
lattice formed by the periods of the holomorphic 1-forms, 
$$\Lambda=\left\{\mathrm{m}+\mathbb{B}\mathrm{n}: 
m,n\in\mathbb{Z}^{g}\right\}.$$ In this section, we discuss how the Abel map 
can be efficiently computed for general points on the Riemann 
surface. In addition, the expansion of the Abel map at a point 
$P\in\mathcal{R}$ near a point $a\in\mathcal{R}$ is 
often needed,
\begin{equation}
    \omega_{j}(P)=\left(U_{a,j}+V_{a,j}k_{a}(P)+\frac{1}{2}W_{a,j}k_{a}(P)^{2}+o(k_{a}(P)^{2})
    \right)dk_{a}(P),\quad j=1,\ldots,g
    \label{abelexp},
\end{equation}
where $k_{a}(P)$ is a local parameter in the vicinity of $a$ 
containing also $P$, and where $U$, $V$, $W$ are vectors in 
$\mathbb{C}^{g}$.

Note that the Abel map is only defined up to periods of the holomorphic
1-forms. Thus we always choose it to be in the fundamental domain given by
\begin{equation}
  \mathrm{q} + \mathbb{B}\,\mathrm{p},\quad \mathrm{p},\mathrm{q}\in \mathbb{R}^{g},
\quad -1/2<p_{i}\leq 1/2, \quad -1/2<q_{i}\leq 1/2, \quad i=1,\ldots,g.
      \label{charabel}
\end{equation}
The Abel map and thus an arbitrary point of the Jacobian 
can be given in terms of the characteristics $(p_{i}, q_{i})$.

\subsection{Point in general position}

For a point $P\in \mathcal{R}$ in general position, i.e., not close to one of
the problem points of the curve (see section \ref{4}) and to infinity, we use
the same approach as for the integrals along the connecting lines in
Fig.~\ref{fig:monpaths}. More precisely if the projection $x(P)$ of $P$ into the
base is such that $|x(P)-b_{i}|>0.8R$ for $i=1,\ldots,N_{c}$ and
$|x(P)|<3\,\mbox{max}_{i=1,\ldots,N_{c}}|b_{i}|$, the point $b_{i_{0}}$ with
minimal distance to $P$ is identified. On the circle of radius $R$ around this
point $b_{i_{0}}$, the collocation point with minimal distance to $P$ is
found. As detailed in Sect.~\ref{5}, the integral of the holomorphic
differentials on a segment of the circle up to this collocation point is
computed by expanding the integrands in terms of Chebychev polynomials and using
(\ref{chebint}). The integral from this collocation point to the point $P$ is
determined as for the integrals along the connecting lines: a number
$N_{l}$ of Chebychev points is introduced in the base on the connecting line
between the projections of the collocation point on the circle and of $P$.  The
line is then lifted to the surface as before by solving (\ref{fdef}) via
\name{roots} and analytically continuing the roots to obtain the integrand on
the connecting line on the surface. Then we use Clenshaw-Curtis integration to
determine the Abel map.

The vectors $U$, $V$ and $W$ in (\ref{abelexp}) are determined in the 
process: $U$ is just the integrand of the Abel map at the last 
collocation point. To obtain $V$ and $W$ respectively, the integrand has to be 
differentiated once respectively twice. Since the integrands are 
quotients of polynomials in $x$ and $y$, this could be done 
essentially by hand. But since the integrand is known already on 
Chebychev collocation points and  since by construction the integrand 
is analytic on the line, we use \emph{Chebychev differentiation 
matrices} as computed in \cite{trefethen1} (see \cite{trefethenweb} 
for a code to generate these matrices). Note that these matrices are 
obtained from the Lagrange polynomial on the collocation points, and 
that this is again a spectral method with an exponential decrease of 
the numerical error with the number of collocation points. Thus the 
derivatives are computed with the same precision as the integral. 

\subsection{Vicinity of finite critical points}

For a point $P$ inside one of the circles around the problem 
points $b_{i_{0}}$,  i.e., with $|x(P)-b_{i_{0}}|\leq 0.8R$, this approach 
will not be efficient. Even if the local coordinate $(x-x_{s})^{1/r}$ 
is used near a branch point $x_{s}$ of order $r$, cancellation errors 
for $x\sim x_{s}$ can affect numerical accuracy. But since the 
monodromies are already known and since  the holomorphic 1-forms are closed, 
the  Cauchy formula (\ref{cauchy}) can 
be also applied to the case that $\mathcal{F}(t)$ is the Abel map. 
More concretely, we consider
\begin{equation}
    \mathcal{F}(x)=\frac{1}{2\pi 
    \mathrm{i}}\int_{\Gamma}^{}\frac{\mathcal{F}(t')dt'}{(x-x_{s})^{1/r}-t'}
    \label{cauchy2},
\end{equation}
where $t'=R^{1/r}\exp(\mathrm{i}\phi/r)$.

This is even a very efficient approach since the holomorphic differentials are
already known on the circle around $b_{i_{0}}$ on Chebychev points. Thus we use
relation (\ref{chebint}) to obtain the Abel map on these collocation points, and
then use (\ref{cauchy}) for the Abel map to obtain its value at point $P$. Since
the Cauchy formula gets numerically problematic if $t'$ is close to $\Gamma$, we
only use it for values of $|x(P)-b_{i_{0}}|\leq 0.8R$. In this case, the Abel map
will be computed up to the precision with which the holomorphic 1-forms are
known on the circle where no cancellation errors are to be expected. The vectors
$U$, $V$ and $W$ in (\ref{abelexp}) appear in the Puiseux expansion of the Abel
map at this point which can be obtained by differentiating (\ref{cauchy2}) with
respect to $t=(x-x_{s})^{1/r}$.  As discussed in section \ref{3}, the involved
integrals are computed to machine precision with this approach.

Note that in this computation, the Abel map to a point in the sheet with the label $k$ is 
always computed starting from the point $A_{k}$, the point in sheet 
$k$ covering the base point. The Abel map between $A_{1}$, which is 
always taken as the base point for the Abel map, and the point 
$A_{k}$ is then added. The connection between these points on the 
surface has been identified by the Tretkoff-Tretkoff algorithm, and 
the integral has been computed already during the monodromy 
computation. Thus this change of the base point comes at virtually no 
computational cost. 

\subsection{Points in the vicinity of infinities}

If the point $P$ is far from all problem points, the Cauchy formula 
will be applied for the circle centered at the origin of radius 
$R_{\infty}$ already used for the Puiseux expansion at infinity in 
section \ref{3}. This circle contains all circles around the critical 
points and touches at least one of them, say $b_{i_{0}}$, in a collocation point. The 
Abel map up to the point where the connecting line to the previous 
critical point on the spanning tree intersects the circle around 
$b_{i_{0}}$ has been already computed. The integral up to the 
collocation point touching the circle with radius $R_{\infty}$ is 
computed as for the period computation with the formula 
(\ref{chebint}). On the circle with radius $R_{\infty}$, the roots 
$y$ have been already determined during the computation of the 
Puiseux expansion at infinity in section \ref{3}. Consequently also 
the holomorphic differentials are known on this circle. Using 
(\ref{chebint}) once more on the circle gives the Abel map on the 
collocation points chosen on this circle.  

With this knowledge, the Cauchy formula (\ref{cauchy2}) can be used 
for the Abel map with $t'=R^{-1/r}\exp(-\mathrm{i}\phi/r)$, 
$\phi\in[0,2\pi]$  and $t=1/x^{1/r}$. The vectors 
$\mathrm{U}$,  $\mathrm{V}$, $\mathrm{W}$ can be obtained as for 
points in the vicinity of finite critical points by differentiating 
(\ref{cauchy2}) with respect to $t$. Again all integrals are computed 
to machine precision.

\subsection{Testing the Abel map}

To test the Abel map, we first consider a hyperelliptic curve of 
genus 2 in Weierstrass normal form,
\begin{equation}
    y^{2}=(x^{2}+1)((x+1)^{2}+1)((x+2)^{2}+1).
    \label{hyperex}
\end{equation}
It is known that the Abel map between branch points of the curve is a 
half period. To test this, we compute the Abel map from the base 
point $A_{1}$ to all branch points and get
\begin{verbatim}
abelmap =

  Columns 1 through 4

   0.9779 - 0.3078i   1.5710 + 0.2243i   1.0710 + 0.2243i   0.4779 - 0.3078i
   1.8651 + 0.8773i   2.3651 + 0.6729i   2.3651 + 0.6729i   2.3651 + 0.8773i

  Columns 5 through 6

   0.9779 - 0.1033i   0.9779 - 0.1033i
   0.9582 + 0.3452i   1.4582 + 0.3452i
\end{verbatim}
corresponding to the Abel map to the branch points $b_{i}$, $i=1,\ldots,6$ in
the order
$-2-\mathrm{i}, -1+\mathrm{i}, \mathrm{i}, -2+\mathrm{i}, -1-\mathrm{i},
-\mathrm{i}$. With
\begin{verbatim}
RieMat =

   1.1862 + 1.0642i   0.0000 - 0.4090i
  -0.0000 - 0.4090i   1.8138 + 1.0642i
\end{verbatim}
we find for 
$\omega(b_{i})-\omega(b_{1})=\mathrm{p}^{i}+\mathbb{B}\mathrm{q}^{i}$, $i=2,\ldots,6$ 
\begin{verbatim}
    q =

    0.5000    0.5000         0         0         0
    0.0000    0.0000         0   -0.5000   -0.5000
\end{verbatim}
and
\begin{verbatim}
    p =
    0        -0.5000   -0.5000   -0.0000   -0.0000
    0.5000    0.5000    0.5000         0    0.5000.
\end{verbatim}
As expected $2\mathrm{p}$, $2\mathrm{q}$ are integers, in fact with an accuracy
of $10^{-16}$. This implies that the Abel map to the branch points is computed
to machine precision.

A further test also used in \cite{RSbookdp} is related to Abel's theorem: recall
that a divisor is a formal symbol $\mathcal{D}=\sum_{i=1}^{K}n_{i}P_{i}$ where
$P_{i}$, $i=1,\ldots,K$ are points on the Riemann surface, and where
$n_{i}\in\mathbb{Z}$. The degree of a divisor is defined as
$\mbox{deg}\mathcal{D}=\sum_{i=1}^{K}n_{i}$.  If the degree of a divisor
vanishes, it is the set of zeros $P_{j}$ (positive $n_{i}$) and poles $Q_{j}$
(negative $n_{i}$) of a meromorphic function on $\mathcal{R}$ (multiplicities
have to be counted). Abel's theorem says that if $\mbox{deg}\mathcal{D}=0$, then
the Abel map of the divisor is a lattice vector in $\Lambda$.

As an example, we consider the function $y$ on the surface defined by
(\ref{cubic}). The Puiseux expansions in section \ref{3} indicate that $y$ has a
pole of order 7 at infinity (there is just one point on the surface covering
infinity) and two zeros of order 3 and 4 respectively covering 0. We get for the
Abel map in the form (\ref{charabel}) ($P_{1}=\infty$, $P_{2}=0^{1}$,
$P_{3}=0^{3}$) (the sheets are denoted by superscripts, $0^{3}$ is 
the point in the third sheet covering $0$)
\begin{verbatim}
[p1,q1]

ans =

    0.5998    0.5615
    0.7193    0.2389

[p2,q2]

ans =

    0.1998    0.7615
    0.1193    0.8389
    
[p3,q3]

ans =

    0.3998    0.1615
    0.9193    0.0389
    \end{verbatim}
The Abel theorem requires that $3p_{2}+4p_{3}-7p_{1}$ and $3q_{2}+4q_{3}-7q_{1}$
are integer vectors. In fact we find for these combinations the vectors
$(2,1)^{T}$ respectively $(1,-1)^{T}$ with a precision of $10^{-9}$.
    
A very efficient test for the Abel map and its derivatives (\ref{abelexp}) is
provided by Fay's trisecant identity \cite{fay} for theta functions. This
identity (\ref{Fay}) for products of theta functions holds for four arbitrary
points on a Riemann surface. Thus it is possible to consider also degenerations
(\ref{Fay1}), (\ref{Fay2}) and (\ref{Fay3}) which are summarized in the
appendix. These identities are very important in the context of
algebro-geometric solutions to integrable equations of which we discuss an
example in the following section. Their purpose here is to allow tests both for
the Abel map and its expansion (\ref{abelexp}).

We consider again the example (\ref{cubic}) and the points $P_{1}=\infty$ 
(there is just one point covering infinity in this example, 
$P_{2}=10^{2}$ (thus a point computed with the Cauchy formula on the 
circle surrounding all branch points), $P_{3}=2^{1}$, a point `in 
general position', and $P_{4}=0^{1}$, a finite singular point. Thus 
this example should test all routines to compute the Abel map. The 
theta functions needed in the various Fay identities are computed to 
machine precision as discussed in section \ref{9}. The Abel map is 
expanded near infinity, a singularity with $r=3$. We find that the 
identity (\ref{Fay}) involving all four points is satisfied to the 
order of $10^{-12}$, the tolerance used in all computations. The same 
holds for the identities (\ref{Fay1}) and (\ref{Fay2}). Identity 
(\ref{Fay3}) is even satisfied to the order $10^{-15}$, but this 
appears to be a coincidence; in other examples it is just satisfied to the order 
of $10^{-10}$ since fourth derivatives of theta functions are 
involved which lowers the maximally achievable accuracy. 

The above tests allow to conclude that the Abel map for a point in 
general position is efficiently computed with the same precision as 
the periods of the surface.

\section{Algebro-geometric solutions to the Kadomtsev-Petviashvili equation}%
\label{10}
As an example we study in this section algebro-geometric solutions to the KP
equation on compact Riemann surfaces. The completely integrable equation can be
seen as $2+1$ dimensional generalization of the celebrated KdV equation. It has
a physical interpretation as describing the propagation of weakly
two-dimensional waves of small amplitude in shallow water as well as similar
physical processes, see for instance \cite{kpsaut} and references therein.  The
KP II equation for the real valued potential $u$ depending on the three real
coordinates $(x,y,t)$ can be written in the form
\begin{equation}
    3u_{yy} + \partial_{x}(6uu_{x}+u_{xxx}-4u_{t})=0
    \label{eq:kp}.
\end{equation}
The KP I equation can be obtained from (\ref{eq:kp}) by changing the sign of the
$u_{xxx}$ term. If $u(x,y,t)$ is a solution to the KP~II equation,
$u(\mathrm{i}x,\mathrm{i}y,\mathrm{i}t)$ formally solves the KP~I equation.
Solutions to KP~I are applicable when surface tension is strong, whereas KP~II
is a good model for weak surface tension.

Algebro-geometric solutions to the KP II equation can be given on an 
arbitrary compact Riemann surface as shown by \cite{K}.
Solutions to the KP equation on the above Riemann surfaces are given 
by the generalization of the Its-Matveev formula for the KdV equation
(see e.g.\ \cite{Mum})
\begin{equation}
    u(x,y,t) = 2\partial_{x}^{2}\ln 
\Theta(\mathrm{U}x+\mathrm{V}y+\mathrm{W}t/2+\mathrm{D})+\frac{c_{1}}{6}
    \label{eq:solution},
\end{equation}
where $\mathrm{D}\in \mathbb{C}^{g}$ and $c_{1}$ is given in (\ref{c1}).  By a
coordinate change $x\to x + tc/3$ one can change a solution $u$ to the KP
equation by $-2c$.  Probably the most elegant way to show that
(\ref{eq:solution}) provides a solution to the KP equation is the use of Fay's
trisecant identity in the form (\ref{Fay3}), see appendix~\ref{sec:Fay}.
Differentiating (\ref{Fay3}) twice, one finds that (\ref{eq:solution}) is indeed
a solution to the KP II equation. KP I solutions can be obtained from
(\ref{eq:solution}) by considering as mentioned
$u(\mathrm{i}x,\mathrm{i}y,\mathrm{i}t)$.

The important question of real and singularity free solutions to the KP equation
was addressed by Dubrovin and Natanzon in \cite{DN}. It was shown that real and
smooth solutions to the KP II equation of the form (\ref{eq:solution}) are
obtained on real Riemann surfaces, i.e., surfaces with all $a_{ij}$ in
(\ref{fdef}) real. Such surfaces have an antiholomorphic involution
$\tau:\mathcal{R}\mapsto \mathcal{R}$, $\tau^{2}=\mbox{id}$, which acts on the
each sheet of the surface as the complex conjugation. The connected components
of the set of fixed points $\mathcal{R}(\mathbb{R})$ of the anti-involution are called the \emph{real
  ovals} of $\tau$.  If $\mathcal{R}\backslash \mathcal{R}(\mathbb{R})$ has two
  components, the surface is called
\emph{dividing}. A curve with the maximal set $g+1$ of real ovals is called an
\emph{M-curve}.  Note that an M-curve is always dividing. In \cite{DN} is was
shown that real, smooth KP II solutions of the form (\ref{eq:solution}) are
obtained on M-curves for $\mathrm{D}\in\mathbb{R}$. Real smooth solutions to the
KP I equation of the form (\ref{eq:solution}) are obtained after the mapping
$x\mapsto \mathrm{i}x$, $y\mapsto \mathrm{i}y$, $t\mapsto \mathrm{i}t$ for
$\mathrm{D}\in \mathrm{i}\mathbb{R}$ on dividing curves.

In \cite{KK1} the numerical evaluation of solutions of the form
(\ref{eq:solution}) was discussed using the example of the Davey-Stewartson and
multi-component NLS equations. An important point in this context is the
introduction of a homology basis adapted to the anti-holomorphic involution
$\tau$ (see \cite{Vin}) for which the regularity conditions can be most easily
identified. Since the Tretkoff-Tretkoff algorithm \cite{tret} implemented in the
code in general does not produce such a basis, a symplectic transformation to
the adapted basis has to be constructed. An algorithm to achieve this was
presented in \cite{KK2} to which the reader is referred for details.  In the
following we will always assume that we work in the adapted basis obtained after
applying this symplectic transformation.

\begin{figure}[htb]
    \centering 
    \includegraphics[width=\textwidth]{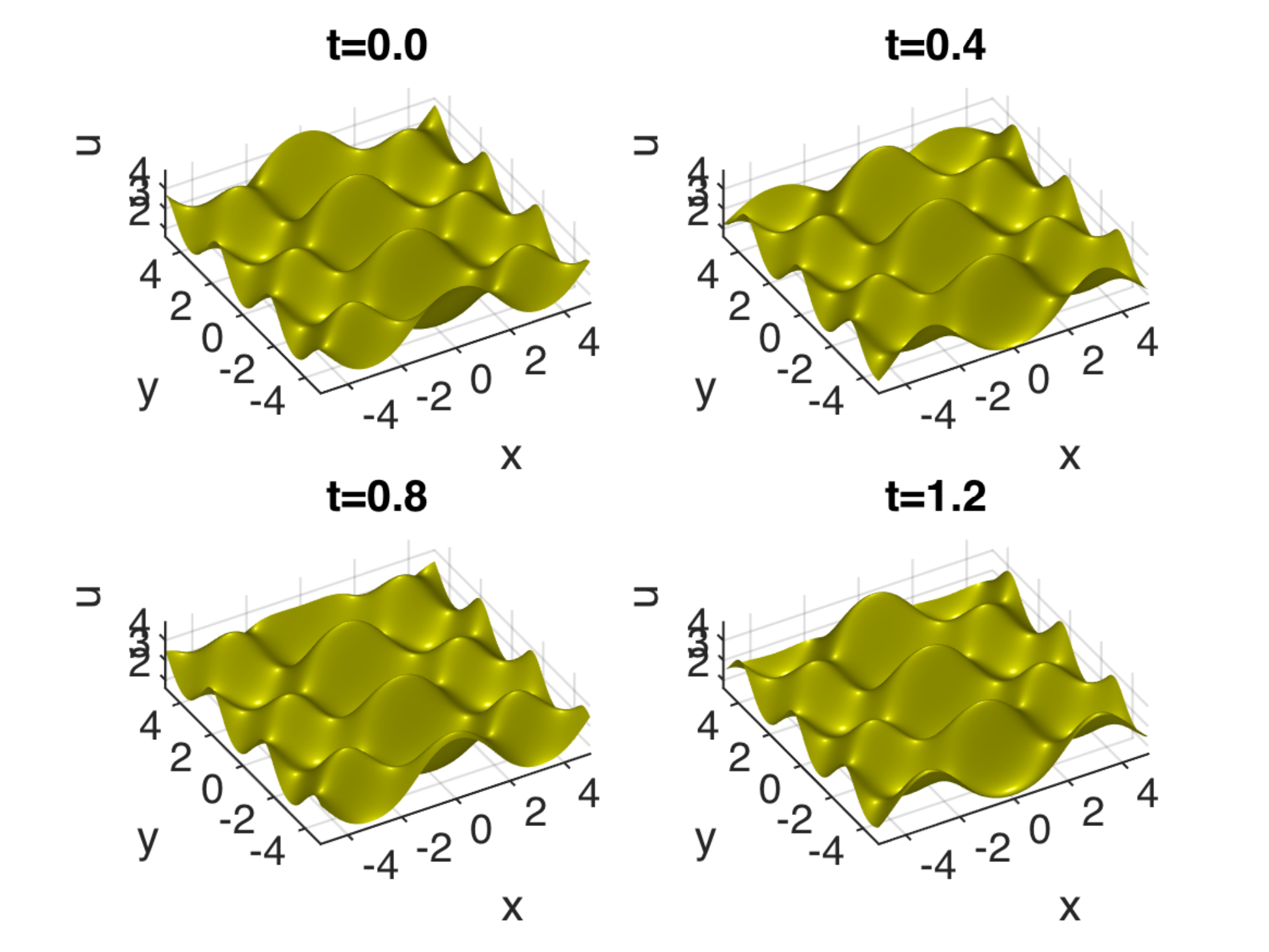}
    \caption{Solution (\ref{eq:solution}) to the KP II equation on 
    the Trott curve (\ref{trott}) for $\mathrm{D}=0$ and 
    $P_{1}=0^{1}$ for different values of $t$. }
    \label{KPtrott}
\end{figure}

Hyperelliptic solutions to the KP equation were discussed in
\cite{lmp}. Therefore we focus here on non-hyperelliptic curves. An example for
a non-hyperelliptic M-curve of genus 3 is the Trott curve \cite{Trott} given by
the algebraic equation
\begin{equation}
    144\,(x^{4}+y^{4})-225\,(x^{2}+y^{2})+350\,x^{2}y^{2}+81=0
    \label{trott};
\end{equation}
it is an M-curve with respect to the anti-holomorphic involution $\tau$ defined
by $\tau(x,y)=(\overline{x},\overline{y})$.  This curve has real branch points
only (and 28 real bitangents, namely, tangents to the curve in two places). In
Fig.\ref{KPtrott} we show the solution (\ref{eq:solution}) to the KP II equation
on the Trott curve for vanishing $\mathrm{D}$ and $P_{1}=0^{1}$. It can be seen that the solution propagates
mainly in the $y$-direction.  The sheets are identified at
\begin{verbatim}
    base =

  -0.9871
  
  ybase =

  -0.0000 + 0.9047i
  -0.0000 - 0.9047i
  -0.1137 + 0.0000i
   0.1137 + 0.0000i
  \end{verbatim}
Note that the accuracy of the solution is controlled pointwise with the twice
differentiated identity (\ref{Fay3}) for theta functions which is identical to
(\ref{eq:kp}). This identity is less well satisfied than for the examples
discussed in previous section for the Abel map since it has been differentiated
twice more (it contains 6th derivatives now, and each derivative leads roughly
to a loss of an order of magnitude in the maximally achievable accuracy). Still
the identity is satisfied in all examples to the order of $10^{-6}$ and better.

\begin{figure}[htb]
    \centering 
    \includegraphics[width=0.7\textwidth]{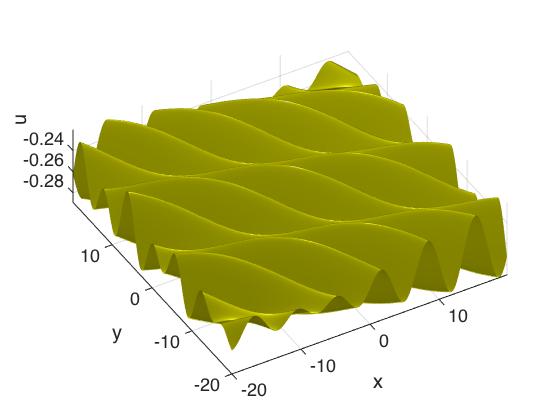}
    \caption{Solution (\ref{eq:solution}) to the KP I equation on 
    the  dividing curve (\ref{divg3}) for $\mathrm{D}=0$ and 
    $P_{1}=0^{1}$ for  $t=0$. }
    \label{KPI3}
\end{figure}

As an example for a dividing curve which is not an M-curve, we consider the
curve given by the equation
\begin{equation}
    30x^4-61x^{3}y+41y^2x^2-43x^2-11y^3x+42xy+y^4-11y^2+9=0
    \label{divg3}
\end{equation}
which was studied in \cite{Dubm} and \cite{Vin}. It is a genus 3 curve, dividing
with respect to the anti-holomorphic involution $\tau$, without real branch
point. This curve admits two real ovals. The KP I solution on this curve for
$\mathrm{D}=0$ and $P_{1}=0^{1}$ for $t=0$ can be seen in Fig.~\ref{KPI3}. The
sheets are identified at
\begin{verbatim}
    base =

  -1.8931 - 0.1931i
  
  ybase =

 -11.1415 - 0.8713i
  -4.5308 - 0.4989i
  -3.9376 - 0.5035i
  -1.2143 - 0.2500i
  \end{verbatim}

\begin{figure}[htb]
    \centering 
    \includegraphics[width=0.7\textwidth]{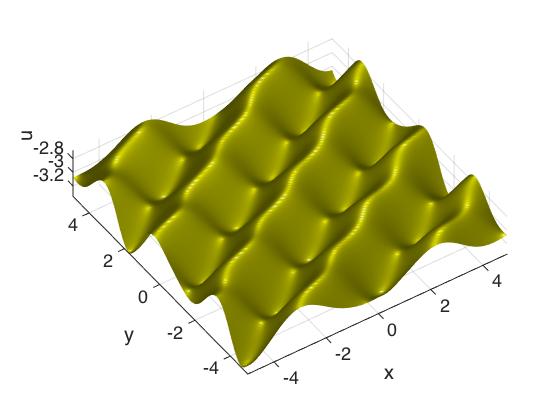}
    \caption{Solution (\ref{eq:solution}) to the KP I equation on 
    the  dividing curve (\ref{divg6}) for $\mathrm{D}=0$ and 
    $P_{1}=0^{1}$ for  $t=0$. }
    \label{KPI6}
\end{figure}

The curve given by the equation 
\begin{equation}
    -180x^5+396yx^4-307x^3y^2+107x^2 y^3+273 x^3-318 x^2 y-17 x 
    y^4+117 x y^2-68 x+y^5-12 y^3+19 y=0
    \label{divg6}
\end{equation}
is a dividing curve of genus 6 without real branch points. We show a 
KP I solution (\ref{divg6}) (after the usual substitution) defined on 
this curve for $\mathrm{D}=0$ and $P_{1}=0^{1}$ in Fig.~\ref{KPI6}. 
The code identifies the sheets as 
\begin{verbatim}
    base =

  -1.8243 - 0.4642i
  
  ybase =

 -11.0742 - 2.7527i
 -10.8111 - 2.0894i
  -4.0601 - 1.2074i
  -3.9298 - 1.2330i
  -1.1382 - 0.6088i
  \end{verbatim}

\section{Outlook}\label{outlook}
In this paper we have outlined a fully numerical method for compact Riemann
surfaces which computes the characteristic quantities of the surface with
spectral accuracy.  An approach to algebraic curves based on finite precision
floating point numbers obviously faces limitations, but as we have shown, these
restrictions are not severe.  We are mainly interested in the study of modular
properties of generic Riemann surfaces of low genus, i.e., of curves which are
regular or do not have singularities corresponding to zeros of very high
multiplicity of the resultant (\ref{resultant}).  In these cases the purely
numerical approach presented here works well and is considerably more efficient
than mixed symbolic-numeric approaches. For algebraic curves of high degree or
with singularities of high order, homotopy tracing methods as in \cite{HS,SW}
are more efficient if accompanied by an \name{endgame} to treat high order zeros
(for instance complex contour integrals as also used in the present paper in a
different context). This is beyond the scope of this paper, but is planned for
future versions of the code.

For an efficient approximation of the theta functions (\ref{theta}) entering the
solutions (\ref{eq:solution}) to the KP equation, it is of course important that
the number of terms in the truncated theta series needed to reach machine
precision is as small as possible. This number is mainly determined by the
eigenvalues of the imaginary part of the Riemann matrix $\mathbb{B}$, see the
discussion in \cite{cam} and \cite{alg1}. Siegel \cite{siegel} has shown that a
symplectic transformation can be found that the diagonal elements of the imaginary
part of $\mathbb{B}$ are always larger than $\sqrt{3}/2$. This does 
not give a priori a lower bound for the eigenvalues of this matrix, but leads 
in general to more equally distributed eigenvalues.  An implementation of
Siegel's algorithm was presented in \cite{deconinck03}. The problem is that
Siegel's method requires that certain matrices appearing during the algorithm
need to be Minkowski ordered which is computationally very expensive. Therefore
the LLL algorithm \cite{LLL} which approximates the Minkowski form and which is
computationally less expensive has been used instead in \cite{deconinck03}, and
is also implemented in the Matlab package presented here. The problem is that
LLL does not provide a controlled approximation of the Minkowski 
conditions. Hence, future versions of the code will contain
an implementation of the Minkowski reduction conditions for small 
genus to test whether this in fact improves the convergence of the 
theta series with respect to the use of the LLL algorithm.

\appendix 
\section{Fay's identities}\label{sec:Fay}
Theta functions satisfy many identities. One of these, Fay's trisecant identity
\cite{fay}, is very interesting and important in the context of completely
integrable equations such as the KP equation studied in this paper. We define
for four arbitrary points $P_{1},P_{2},P_{3},P_{4}\in\mathcal{R}$ the cross
ratio function ($\Theta^{*}$ is a theta function with a nonsingular half-integer
characteristic)
\begin{equation}
    \lambda_{1234}=\frac{\Theta^{*}(\omega(P_{1})-\omega(P_{2}))\Theta^{*}(\omega(P_{3})-\omega(P_{4}))}{
    \Theta^{*}(\omega(P_{1})-\omega(P_{4}))\Theta^{*}(\omega(P_{3})-\omega(P_{2}))}
    \label{eq:cross3}\;,
\end{equation}
which is a function on $\mathcal{R}$ that vanishes for $P_{1}=P_{2}$
and $P_{3}=P_{4}$ and has poles for $P_{1}=P_{4}$ and $P_{2}=P_{3}$.
Then the following
    identity holds:
    \begin{equation}
      \begin{aligned}
        \lambda_{3124}\,\Theta(\mathrm{z}+\smallint_{P_{2}}^{P_{3}})\,
        &\Theta(\mathrm{z}+\smallint_{P_{1}}^{P_{4}}) +\lambda_{3214}\,
        \Theta(\mathrm{z}+ \smallint_{P_{1}}^{P_{3}})\,\Theta(\mathrm{z}+
        \smallint_{P_{2}}^{P_{4}})\\
       & =\Theta(\mathrm{z})\;\Theta(\mathrm{z}+\smallint_{P_{2}}^{P_{3}}+
        \smallint_{P_{1}}^{P_{4}})\;,
      \end{aligned}
    \label{Fay}
\end{equation} 
   where $\mathrm{z}\in \mathbb{C}^{g}$  and
   $\smallint_{P}^{Q}=\omega(Q)-\omega(P)$. The 
   integration paths in (\ref{Fay}) have to be chosen in a way not to 
   intersect the canonical cycles.

Degenerate versions  of  Fay's 
identity lead to identities for derivatives of 
theta functions. Let $k(P)$ be a 
local coordinate near $P$. 
We denote by $D_P$ the operator for the directional derivative 
along the basis of holomorphic differentials, 
acting on theta functions, and similarly $D_{P}'$ and $D_{P}''$ 
the directional derivatives along $\mathrm{V}$ and $\mathrm{W}$,
\begin{eqnarray} 
D_P\Theta_{\mathrm{p}\mathrm{q}}(\mathrm{z})&=&\langle \nabla 
\Theta_{\mathrm{p}\mathrm{q}}(\mathrm{z}),\mathrm{U}\rangle\;, 
\nonumber\\ 
D_P'\Theta_{\mathrm{p}\mathrm{q}}(\mathrm{z})&=&\langle \nabla 
\Theta_{\mathrm{p}\mathrm{q}}(\mathrm{z}),
\mathrm{V}\rangle\;, \nonumber\\
D_P''\Theta_{\mathrm{p}\mathrm{q}}(\mathrm{z})&=&\langle \nabla 
\Theta_{\mathrm{p}\mathrm{q}}(\mathrm{z}), 
\mathrm{W}\rangle\;.\label{eq:defD}
\end{eqnarray}

In the limit $P_{4}\to P_{2}$, one finds for (\ref{Fay})
\begin{equation}
    D_{P_{2}}\ln \frac{\Theta(\mathrm{z}+
    \smallint_{P_{1}}^{P_{3}})}{\Theta(\mathrm{z})}=p_{1}(P_{1},P_{2},P_{3})+p_{2}(P_{1},P_{2},P_{3})
    \frac{\Theta(\mathrm{z}+
    \smallint_{P_{2}}^{P_{3}})\Theta(\mathrm{z}+
    \smallint_{P_{1}}^{P_{2}})}{\Theta(\mathrm{z}+
    \smallint_{P_{1}}^{P_{3}})\Theta(\mathrm{z})}
    \label{Fay1},
\end{equation}
where 
\begin{align}
    p_{1}(P_{1},P_{2},P_{3}) &=  D_{P_{2}}\ln \frac{\Theta^{*}(
    \smallint_{P_{1}}^{P_{2}})}{\Theta^{*}(\smallint_{P_{3}}^{P_{2}})},
    \nonumber\\
    p_{2}(P_{1},P_{2},P_{3}) & =\frac{\Theta^{*}(
    \smallint_{P_{1}}^{P_{3}})D_{P_{2}}\Theta^{*}(0)}{\Theta^{*}(
    \smallint_{P_{2}}^{P_{3}})\Theta^{*}(
    \smallint_{P_{2}}^{P_{1}})}
    \label{pi}.
\end{align}

In the limit $P_{3}\to P_{1}$, equation (\ref{Fay1}) yields
\begin{equation}
    D_{P_{1}}D_{P_{2}}\ln \Theta(\mathrm{z})=q_{1}(P_{1},P_{2})+q_{2}(P_{1},P_{2})
    \frac{\Theta(\mathrm{z}+
    \smallint_{P_{2}}^{P_{1}})\Theta(\mathrm{z}+
    \smallint_{P_{1}}^{P_{2}})}{\Theta(\mathrm{z})^{2}}
    \label{Fay2},
\end{equation}
where 
\begin{align}
    q_{1}(P_{1},P_{2}) &=  D_{P_{1}}D_{P_{2}}\ln \Theta^{*}(
    \smallint_{P_{1}}^{P_{2}}),
    \nonumber\\
    q_{2}(P_{1},P_{2}) & =\frac{D_{P_{1}}\Theta^{*}(0)D_{P_{2}}\Theta^{*}(0)}{\Theta^{*}(
    \smallint_{P_{1}}^{P_{2}})^{2}}
    \label{qi}.
\end{align}

In the limit $P_{2}\to P_{1}$, equation (\ref{Fay2}) can be cast into 
the form (we suppress the index $P_{1}$ at the derivatives)
\begin{equation}
    D^{4}\ln\Theta(\mathrm{z})+6(D^{2}\ln 
    \Theta(\mathrm{z}))^{2}+3D'D'\ln \Theta(\mathrm{z})-2DD''\ln 
    \Theta(\mathrm{z})+c_{1}D^{2}\ln \Theta(\mathrm{z}) +c_{2}=0
    \label{Fay3},
\end{equation}
where 
\begin{equation}
    c_{1}=2\frac{D''\Theta^{*}(0)}{\Theta^{*}(0)}-4\frac{D^{3}\Theta^{*}(0)}{D\Theta^{*}(0)}
    -3\left(\frac{D'\Theta^{*}(0)}{D\Theta^{*}(0)}\right)^{2};
    \label{c1}
\end{equation}
the constant $c_{2}$ is not needed here (its somewhat involved 
expression can be obtained by expanding $q_{1}$ and $q_{2}$ 
(\ref{qi}) in the considered limit to fourth order in the local 
parameter near $P_{1}$).


\end{document}